\documentstyle[12pt]{article}

\newcommand{\beq}[1]{\begin{equation} \label{#1} }
\newcommand{\eeq}    {\end{equation}}
\newcommand{\tst}{\textstyle \mathstrut}
\newcommand{\dst}{\displaystyle}
\newcommand{\rdown}[1]{\mathop{#1}\nolimits_{|\,r>R}}
\newcommand{\arctg}{\mathop{\rm arctg}\nolimits}
\newcommand{\al}{\alpha}
\newcommand{\ep}{\epsilon}
\newcommand{\ve}{\varepsilon}
\newcommand{\kp}{\mbox{\ae}}
\newcommand{\lal}{\lambda_l}
\newcommand{\om}{\omega}
\newcommand{\si}{\sigma}
\newcommand{\sig}{\mbox{\boldmath $\sigma$}}
\newcommand{\oh}{\frac{1}{2}}
\newcommand{\th}{\frac{3}{2}}
\newcommand{\toh}{\frac{\textstyle 1}{\textstyle 2}}
\newcommand{\tth}{\frac{\textstyle 3}{\textstyle 2}}
\newcommand{\pr}{\mbox{$\scriptstyle \,\prime$}}
\newcommand{\vph}
{\vphantom{\displaystyle \left(\frac{I}{I}
\right)^{\tst (3)}_{\tst (3)}}}

\begin{document}
\pagenumbering{roman}

\mbox{}
\vspace{1cm}

\noindent{\large Russian Research Center "Kurchatov
Institute"}
\vspace{1cm}

\begin{tabbing}
{\large A.L.Barabanov\footnotemark{}}
 \` {\large Preprint IAE-5922/2}\\
\end{tabbing}\footnotetext{\normalsize e-mail: x1052@kiae.su}
\vspace{3cm}

\noindent{\large MODEL FOR RESONANCE ENHANCEMENT\\[\medskipamount]
OF P- AND T-NONINVARIANT EFFECTS\\[\medskipamount]
IN NEUTRON REACTIONS}
\vspace{4cm}

\noindent {\large Submitted to Nucl.Phys.A}
\vspace{4cm}

\centerline{\large Moscow -- 1995}

\newpage

\noindent UDK 539.17
\vspace{1cm}

\noindent Key words: neutron, nucleus, resonance,
parity violation, time reversal
invariance, coupled channels
\vspace{2cm}

\noindent We consider a simplified model for resonant
neutron-nucleus
interaction with coupled channels. An analytical solution is given
for two coupled channels and arbitrary neutron orbital momentum. A
case of a week channel coupling, corresponding to narrow
Breit-Wigner resonance, is analyzed in details. As far as the
total width of a resonance coincides with the neutron width, the
model is directly appropriate only for light nuclei. We study a
mixing of two narrow s- and p-wave resonances by P- and
P-,T-noninvariant potentials in the first order of perturbation
theory. As an example a close-lying pair of s- and p-wave
resonances of the $^{35}$Cl nucleus is considered. In a resonance
with an orbital momentum $l$ an enhancement of mixing amplitude in
comparison with potential scattering is
$\sim(\omega/\gamma_l)^{1/2}(1/(kR)^{2l+1})$, where $\omega$ is a
characteristic single-particle energy and $\gamma_l$ is a reduced
width of a resonance. The favorable possibilities are shown to
exist on thick targets for measurements beyond the resonance
widths. In particular, an interference minimum near s-wave
resonance is of interest for P-odd neutron spin rotation on light
spinless nuclei.
\vspace{6cm}

\noindent \copyright\ Russian Research Center
"Kurchatov Institute", 1995

\newpage
\pagenumbering{arabic}

\section{Introduction}
\label{s1}

The question of scale and nature of T-invariance violation in
fundamental interactions remains open. So far only the results,
obtained in K$^0$-meson decays \cite{1a}, evidence for such
violation. Many researches of T-noninvariant effects in nuclear
processes, in neutron $\beta$-decay, as well as studies of
electric dipole moments of elementary particles have lowered the
upper limit on mixing of T-noninvariant forces to T-invariant ones
to the level of $10^{-3}-10^{-4}$ (see, e.g., Ref.\cite{2a}).
Now possible T-noninvariant effects in isolated
compound-resonances are widely discussed \cite{3a}-\cite{3h}. An
interest to them results from huge enhancement of P-noninvariant
effects. As it was shown for the first time in Ref.\cite{4a},
transmission asymmetry of neutrons with opposite helicities in
p-wave resonances may be as much as $10^{-3}-10^{-1}$ that is 5--7
order greater than a similar effect in N--N scattering. The review
of recent developments in the study of P-invariance violation in
neutron resonances is given in Ref.\cite{4b}. Now a sign
correlation of P-odd effects, measured in several resonances of
the $^{232}$Th nucleus \cite{10a}, attracts particular attention.
This result is discussed, e.g., in Refs.\cite{10b}-\cite{10f}.

Usually one distinguishes dynamic and structural (or kinematic)
enhancements. The first is due to high density of resonances
\cite{5a}-\cite{5c}. As it was shown in Refs.\cite{9a}-\cite{9c},
this enhancement should take place for T-noninvariant effects as
well as for P-noninvariant ones. A structural enhancement
\cite{6a} (see, also, \cite{7a}) arises
in p-wave resonances, while in
s-wave ones we have a structural suppression. Both type of
enhancement do not represent the fact that the observables peak in
resonances. So a concept of resonance enhancement was brought into
practice in Refs.\cite{8a,8b}, where an energy dependence of
P- and T-noninvariant effects near s- and p-wave resonances has
been investigated. This type of enhancement results from an
increase of
time spent by a neutron in the weak-interaction field of a target.
It was
first mentioned in Refs.\cite{11a,11b}.

As far as a structure of highly excited nuclei is very complex,
there is no consistent microscopic theory of neutron resonances.
Thus phenomenological models as R-matrix theory
\cite{13a} or shell model approach to nuclear reactions \cite{14a}
are used to estimate the observables. There exist, however, some
uncertainties in these models in the phases of the S-matrix
elements, caused by potential scattering or contributions
of distant resonances. This may lead to uncontrolled fake effects,
which make difficult a search of T-invariance violation
(see, e.g., Ref.\cite{3b}).

In this paper we study P- and T-noninvariant effects in
simplified,
but exactly solvable model of neutron-nucleus interaction. The
model reproduces narrow resonances, occurring owing to
excitations of selected degrees of freedom of a target. The basis
for the model is a scheme of coupled channels \cite{15a},
described in section \ref{s2}. In section \ref{s3}
P- and T-noninvariant observables are expressed in terms of
S-matrix
elements. A method for calculation of P- and T-noninvariant
corrections to S-matrix is presented in section \ref{s4}. The
exactly
solvable model of resonant neutron-nucleus interaction is stated
in section \ref{s5}. In section \ref{s6} this model is used for
description of a close-lying pair of s- and p-wave resonances
and their mixing by P- and T-noninvariant interactions. The
results of illustrative calculation of P- and T-noninvariant
observables are presented
in section \ref{s7}. In section \ref{s8} a summary of the most
important conclusions is given.

\section{Scheme of coupled channels}
\label{s2}

We consider an interaction of a neutron with a nucleus of mass A.
Let
$\hat{H}_A$ is a nuclear hamiltonian, and $\ve_{\al}$ and
$\psi_{\al}(\tau)$ are its eigenvalues and orthonormal
eigenfunctions
\beq{2.1}
\hat{H}_A\psi_{\al}(\tau)=\ve_{\al}\psi_{\al}(\tau),
\qquad
<\psi_{\al'}|\psi_{\al}>=\delta_{\al'\al},
\eeq
$\tau$ is a set of internal variables. The hamiltonian $\hat{H}_A$
will be considered as invariant with respect to rotations and
space and time inversions (R-, P- and T-invariant, respectively).
We assume that the nuclear spectrum is purely discrete. An index
$\al$
includes spin $I$, its projection $\mu$ on an axis z, parity $\pi$
and number $i$, which distinguishes states with the same $I$,
$\mu$
and $\pi$. An energy $\ve_{\al}$ does not depend on $\mu$ because
of R-invariance of hamiltonian $\hat{H}_A$

In a center-of-mass system of neutron and nucleus a total
hamiltonian is of the form
\beq{2.2}
\hat{H}=-\frac{\hbar^2}{2m}
\frac{\partial^2}{\partial{\bf r}^2}+
\hat{U}+\hat{H}_A,
\eeq
where ${\bf r}={\bf r}_n-{\bf R}_A$ is a neutron radius-vector
with respect to a target center of mass, $m$ is a reduced mass,
and $\hat{U}$ is an operator of interaction of a neutron with
target
nucleons.

Let us introduce the orthonormal single-particle spin-angular
functions
\beq{2.3}
|j\nu;l>=\sum_{m\si}C^{j\nu}_{lms\si}i^l
Y_{lm}({\bf r})\chi_{s\si},
\qquad
<j'\nu';l'|j\nu;l>=
\delta_{j'j}\delta_{\nu'\nu}\delta_{l'l}.
\eeq
Here $\chi_{s\si}$ is a spinor, describing a state of a neutron
with a projection $\si$ of a spin $s=1/2$ on an axis z,
$Y_{lm}({\bf r})$
are spherical harmonics, $l$ and $m$ are relative neutron-nucleus
orbital momentum and its projection on an axis z, and
$C^{j\nu}_{lms\si}$ are Clebsh-Gordan coefficients. Angular
momentum ${\bf j}={\bf l}+{\bf s}$ will be named a neutron total
angular momentum. Functions
\beq{2.4}
|JM;lj\al>=\sum_{\nu\mu}C^{JM}_{j\nu I\mu}
|j\nu;l>\psi_{\al}(\tau)
\eeq
are also orthonormal
\beq{2.5}
<J'M';l'j'\al'|JM;lj\al>=
\delta_{J'J}\delta_{M'M}
\delta_{l'l}\delta_{j'j}\delta_{\al'\al}.
\eeq
Here $J$ and $M$ are a total angular momentum and its projection
on an axis z. The phases are so chosen that the wave functions
have usual transform properties with respect to time inversion
\cite{16a}
\begin{eqnarray}
\lefteqn{\hat{T}\psi_{I\mu}(\tau)=
(-1)^{I+\mu}\psi_{I-\mu}(\tau),}&
\phantom{\hat{T}\psi_{I\mu}(\tau)=
(-1)^{I+\mu}\psi_{I-\mu}(\tau),}&
\hat{T}\chi_{s\si}=(-1)^{s+\si}\chi_{s-\si},
\nonumber\\[\medskipamount]
\lefteqn{\hat{T}|j\nu;l>=(-1)^{j+\nu}|j-\nu;l>,}&
\phantom{\hat{T}|j\nu;l>=(-1)^{j+\nu}|j-\nu;l>,}&
\hat{T}|JM;lj\al>=(-1)^{J+M}|J-M;lj\al>.
\label{2.6}
\end{eqnarray}

Let us present a total wave function of neutron and nucleus as a
series in state vectors (\ref{2.4})
\beq{2.7}
\Psi({\bf r},\tau)=\sum_{JM}\sum_{lj\al}
R^{(\al)}_{ljJM}(r)|JM;lj\al>.
\eeq
Substituting this expansion in the Schr\"odinger equation
\beq{2.8}
\hat{H}\Psi({\bf r},\tau)=E\Psi({\bf r},\tau),
\eeq
and projecting on $<JM;lj\al|$, we get
$$
\frac{d^2R^{(\al)}_{ljJM}(r)}{dr^2}+
\frac{2}{r}\frac{dR^{(\al)}_{ljJM}(r)}{dr}-
\frac{l(l+1)}{r^2}R^{(\al)}_{ljJM}(r)+
\frac{2m(E-\ve_{\al})}{\hbar^2}R^{(\al)}_{ljJM}(r)-{}
$$
\beq{2.9}
{}-\frac{2m}{\hbar^2}\sum_{l'j'\al'}
<JM;lj\al|\hat{U}|JM;l'j'\al'>
R^{(\al')}_{l'j'JM}(r)=0.
\eeq
The matrix element of an operator $\hat{U}$ is diagonal on $J$ and
$M$ and does not depend on $M$ because of R-invariance of
interaction.

We assume that the interaction $\hat{U}$ is short-range and
vanishes at $r>R$. Then outside the interaction region the total
wave function is an eigenfunction of hamiltonian of free motion.
So it can be presented as an expansion in spherical Hankel
functions
$h^{(\pm)}_l(x)$
\begin{eqnarray*}
\rdown{\Psi_{\al_0}({\bf r},\tau)}&=&
\sum_{JM}\sum_{lj}a(ljJM)
\Bigl(h^{(-)}_l(k_{\al_0}r)|JM;lj\al_0>+{}
\\[\medskipamount]
&&\phantom{\sum_{JM}\sum_{lj}}+
\sum_{l'j'}S_J(lj\al_0 \to l'j'\al_0)
h^{(+)}_{l'}(k_{\al_0}r)|JM;l'j'\al_0>\Bigr)+{}
\end{eqnarray*}
\beq{2.10}
{}+\sum_{\al \ne \al_0}
\left(\frac{k_{\al}}{k_{\al_0}}\right)^{1/2}
\sum_{JM}\sum_{lj}a(ljJM)
\sum_{l'j'}S_J(lj\al_0 \to l'j'\al)
h^{(+)}_{l'}(k_{\al}r)|JM;l'j'\al>.
\eeq
This function describes a neutron scattering by nucleus being in
the state
$\al_0$. A relative momentum, $\hbar k_{\al_0}$, in an
entrance channel is given by equation
\beq{2.11}
E=\frac{(\hbar k_{\al_0})^2}{2m}+\ve_{\al_0}.
\eeq
Elements of S-matrix, $S_J(lj\al_0\to l'j'\al_0)$, correspond to
elastic scattering with a total angular momentum $J$. Orbital,
$l$, and total, $j$, neutron angular momenta can change in elastic
scattering within the limits of the rules of angular momentum
summation
(\mbox{$|l-s|\le j\le l+s$}, \mbox{$|I-j|\le J\le I+j$}). Nuclear
transition to a state $\al\ne\al_0$ corresponds to inelastic
scattering. A relative momentum in an inelastic channel is
\beq{2.12}
\hbar k_{\al}=\left(2m(E-\ve_{\al})\right)^{1/2}.
\eeq
If $\ve_{\al}>E$, then $\hbar k_{\al}=i(\hbar q_{\al})$, where
$q_{\al}$ is a real positive quantity. Taking into account an
asymptotic of Hankel functions
\beq{2.13}
h^{(\pm)}_l(x)
\mathrel{\mathop{\longrightarrow}\limits_{x\to +\infty}}
(\mp i)^{l+1}\frac{\exp(\pm ix)}{x},
\eeq
we see, that the channel with the energy $\ve_{\al}>E$ is closed
(the wave function falls off exponentially at long range). The
function
(\ref{2.10}) contains factors
\beq{2.14}
a(ljJM)=2\pi\sum_{\nu\mu m\si}
C^{JM}_{j\nu I\mu}C^{j\nu}_{lms\si}
a_{\mu}(I)a_{\si}(s)Y^*_{lm}({\bf k}_{\al_0}),
\eeq
where $a_{\mu}(I)$ and $a_{\si}(s)$ are amplitudes of nucleus and
neutron states with spin projections $\mu$ and $\si$ on an axis z,
respectively.

Comparing a general expansion (\ref{2.7}) for the total wave
function with its form (\ref{2.10}) outside the interaction
region, we find for radial functions
\begin{eqnarray}
\rdown{R^{(\al)}_{ljJM}(r)}&=&
\left(\frac{k_{\al}}{k_{\al_0}}\right)^{1/2}
\sum_{l_0j_0}a(l_0j_0JM)
\Bigl(\delta_{ll_0}\delta_{jj_0}\delta_{\al \al_0}
h^{(-)}_l(k_{\al}r)+{}\nonumber
\\[\medskipamount]
&&\phantom{\left(\frac{k_{\al}}{k_{\al_0}}\right)^{1/2}}+
S_J(l_0j_0\al_0\to lj\al)h^{(+)}_l(k_{\al}r)\Bigr).
\label{2.15}
\end{eqnarray}
Let us write these radial functions in the form
\beq{2.16}
R^{(\al)}_{ljJM}(r)=2\sum_{l_0j_0}
a(l_0j_0JM)\frac{F^J_{n_0n}(r)}{r},
\eeq
where an index $n$ includes $l$, $j$ and $\al$. An index $n_0$
suggests, that the functions $F^J_{n_0n}(r)$ describe
scattering of a neutron with initial orbital, $l_0$, and total,
$j_0$, angular momenta on a nucleus, being in a state $\al_0$.
Substituting the functions (\ref{2.16}) in the equations
(\ref{2.9}) and separating the terms at the same factors
$a(l_0j_0JM)$, we get
\beq{2.17}
\frac{d^2F^J_{n_0n}(r)}{dr^2}-
\sum_{n'}r<JM;n|\frac{2m\hat{U}}{\hbar^2}+
\frac{l(l+1)}{r^2}-k^2_n|JM;n'>
\frac{F^J_{n_0n'}(r)}{r}=0.
\eeq
Outside the interaction region the functions $F^J_{n_0n}(r)$ take
the form
\beq{2.18}
\rdown{F^J_{n_0n}(r)}=
\frac{1}{2(k_{n_0}k_n)^{1/2}}
\left(\delta_{nn_0}\,k_nr\,h^{(-)}_l(k_nr)+
S_J(n_0\to n)\,k_nr\,h^{(+)}_l(k_nr)\right).
\eeq
We assume that the interaction has no singularity at the point
$r=0$. Then the radial functions should be regular at the origin
\beq{2.19}
F^J_{n_0n}(0)=0.
\eeq

Thus, in the scheme of coupled channels, that we have considered,
the
set of equations (\ref{2.17}) completely determines the dynamics
of nuclear reaction. Solving these equations with boundary
conditions (\ref{2.18}), (\ref{2.19}), we get the S-matrix
elements and, consequently, all observables.

\section {P- and T-noninvariant observables}
\label{s3}

In neutron transmission experiments one studies asymmetry of a
total cross section and spin rotation, both caused by parity
nonconservation or possible T-invariance violation. The
observables may be expressed in terms of a forward elastic
scattering amplitude. This amplitude, averaged over spin states of
nuclei, is a matrix $2\times 2$ in a neutron spin space. Thus
it can be presented as an expansion in Pauli matrixes $\si_i$
\beq{3.1}
f(0)=F_0+(\sig {\bf n})F_1,
\eeq
where ${\bf n}$ is some unit vector.

According to optical theorem a total cross section is
\beq{3.2}
\si_t=\frac{4\pi}{k}\left({\rm Im}F_0+
p_1(s)({\bf n}_s{\bf n})\,{\rm Im}F_1\right).
\eeq
Here $p_1(s)$ is a neutron polarization,
and ${\bf n}_s$ is a unit vector along an axis of polarization. We
see, that an asymmetry of a total cross section for neutrons,
polarized along a direction ${\bf n}$ or opposite to it, is
expressed in terms of imaginary part of factor $F_1$. On the other
hand, using the methods of neutron optics \cite{17a}, we obtain
for polarization of neutrons, passed through a target of density
$\rho$ and thickness $d$
\beq{3.3}
p'_1(s){\bf n}'_s=p_1(s){\bf n}_s+
\frac{4\pi n}{k}p_1(s)[{\bf n}_s{\bf n}]\,{\rm Re}F_1-
\frac{4\pi n}{k}{\bf n}\,{\rm Im}F_1,
\eeq
where $n=\rho\,d$ is a number of nuclei on a unit of target area.
Clearly, an angle of spin rotation around a vector ${\bf n}$ is
determined by real part of factor $F_1$. The last term in
Eq.(\ref{3.3}) describes polarization of passed neutrons, arising
owing to asymmetry of a total cross section.

If target nuclei are not oriented, a vector ${\bf n}$ in
Eq.(\ref{3.1}) coincides with a unit vector ${\bf n}_k$ along a
momentum $\hbar{\bf k}$ of incident neutrons. A relevant
P-noninvariant correlation $(\sig{\bf n}_k)$ was considered for
the first time in Ref.\cite{18a}. As noted in introduction, this
correlation was being studied intensively last years. It was
pointed out in Refs.\cite{19a,19b} that nuclear polarization
makes possible the study of P- and T-noninvariant correlation
$(\sig[{\bf n}_k{\bf n}_I])$, where ${\bf n}_I$ is a  unit vector
along an axis of nuclear orientation.
It was shown in Refs.\cite{20a}-\cite{20c} that T-noninvariant,
but P-even correlation
$(\sig [{\bf n}_k{\bf n}_I])({\bf n}_k{\bf n}_I)$ can be studied
on an aligned target. In this paper we restrict our attention to
P- and P-,T-noninvariant effects, so we shall consider only
a polarized target.

Let $p_1(I)$ is nuclear polarization. Taking into account only
s- and p-waves, we express the factors $F_0$ and $F_1$ in terms of
the S-matrix elements, $S_J(lj\to l'j')$, corresponding to elastic
scattering (we omit everywhere in this section an elastic-channel
index $\al_0$)
\beq{3.4}
F_0=a_0+b_2p_1(I)({\bf n}_k{\bf n}_I),
\eeq
\beq{3.5}
F_1{\bf n}=a_1p_1(I){\bf n}_I+
a_2p_1(I)(3{\bf n}_k({\bf n}_I{\bf n}_k)-{\bf n}_I)+
b_1{\bf n}_k+c_1p_1(I)[{\bf n}_k{\bf n}_I].
\eeq
A coefficient $a_0$ takes the form
\beq{3.6}
a_0=\frac{i}{2k}\sum_Jg_J
\sum_{lj}(1-S_J(lj\to lj)),
\eeq
where $g_J=(2J+1)/2(2I+1)$ is a statistical factor. Coefficients
$a_1$ and $a_2$ specify spin-spin interaction of a neutron and
nucleus
\beq{3.7}
a_1=\frac{i}{2k}\sum_J\Bigl(
A_J^{(1)}(1-S_J(0\oh \to 0\oh))+
\sum_{jj'}A_{Jjj'}^{(2)}
(\delta_{jj'}-S_J(1j\to 1j'))\Bigr),
\eeq
\beq{3.8}
a_2=\frac{i}{2k}\sum_{Jjj'}A_{Jjj'}^{(3)}
(\delta_{jj'}-S_J(1j\to 1j')).
\eeq
Explicit expressions for numerical factors $A$, as well as for
factors $B$ and $C$, defined below, are presented in an appendix.
Coefficients $b_1$ and $b_2$ for P-odd correlations are given
by formulas
\beq{3.9}
b_1=\frac{i}{2k}\sum_Jg_J
(S_J(0\oh \to 1\oh)+S_J(1\oh \to 0\oh)),
\eeq
\beq{3.10}
b_2=\frac{i}{2k}\sum_{Jj}B_{Jj}
(S_J(0\oh \to 1j)+S_J(1j \to 0\oh)).
\eeq
Finally, we have for magnitude of possible P-,T-noninvariant
effect
\beq{3.11}
c_1=\frac{1}{2k}\sum_{Jj}C_{Jj}
(S_J(0\oh \to 1j)-S_J(1j \to 0\oh)).
\eeq
We see, the coefficients $b_1$, $b_2$ and $c_1$ are due to
transitions from s- to p-wave and vice versa, that is possible
only
at parity nonconservation. While the coefficient $c_1$ differs
from zero, if additionally a symmetry of S-matrix with respect to
the main diagonal is violated.

If a target is not polarized, then $p_1(I)=0$, therefore
$F_0=a_0$, $F_1=b_1$, ${\bf n}={\bf n}_k$. Thus, as noted, the
unique correlation $(\sig {\bf n}_k)$ remains. Let $\si_+$ and
$\si_-$ are total cross section for neutrons, completely polarized
along the vector ${\bf n}_k$ and opposite to it, respectively.
Thus, we obtain for P-noninvariant asymmetry of a total cross
section
\beq{3.12}
\Delta\si_P\equiv\si_+-\si_-=
\frac{4\pi}{k^2}\sum_Jg_J
{\rm Re}(S_J(0\oh \to 1\oh)+S_J(1\oh \to 0\oh)).
\eeq
At the same time according to Eq.(\ref{3.3}) the angle of spin
rotation for transversely polarized neutrons, counted in a
direction of right-hand rotation
along the vector ${\bf n}_k$, equals
\beq{3.13}
\chi_P=\frac{2\pi n}{k^2}\sum_Jg_J
{\rm Im}(S_J(0\oh \to 1\oh)+S_J(1\oh \to 0\oh)).
\eeq

We turn now to the case, when target nuclei are polarized
transversely the neutron momentum (${\bf n}_I\perp {\bf n}_k$).
Let $\si_{\uparrow}$ and $\si_{\downarrow}$ are total cross
section for neutrons, completely polarized along the vector
$[{\bf n}_k{\bf n}_I]$ and opposite to it, respectively. Then we
have for P-,T-noninvariant asymmetry of a total cross section
\beq{3.14}
\Delta\si_{PT}\equiv\si_{\uparrow}-\si_{\downarrow}=
\frac{4\pi}{k^2}\sum_JC_{Jj}
{\rm Im}(S_J(0\oh \to 1j)-S_J(1j \to 0\oh)).
\eeq
A contribution of P-,T-noninvariant correlation to the angle of
spin rotation around the vector $[{\bf n}_k{\bf n}_I]$ for
neutrons, initially polarized along ${\bf n}_I$, is
\beq{3.15}
\chi_{PT}=-\frac{2\pi n}{k^2}\sum_JC_{Jj}
{\rm Re}(S_J(0\oh \to 1j)-S_J(1j \to 0\oh)).
\eeq
In reality spin-spin forces, as well as P-odd
correlation $(\sig {\bf n}_k)$ make difficult a measurement
of the asymmetry (\ref{3.14}) and the angle (\ref{3.15}). Now the
ways of suppression of the fake effects are intensively discussed
\cite{21a}-\cite{21c}.

In transmission experiments an asymmetry
\beq{3.16}
\ep=\frac{N^+-N^-}{N^++N^-},
\eeq
is directly measured, where $N^+$ and $N^-$ are numbers of
neutrons, passed through a target for two opposite values of their
initial polarization. It is easy to show (see, e.g.,
Ref.\cite{4a}) that these observable asymmetries express by
asymmetries of a total cross section in the following way
\beq{3.17}
\ep_{P(PT)}=-n\,\frac{\Delta \si_{P(PT)}}{2}.
\eeq

\section {First order corrections to S-matrix}
\label{s4}

Weak interaction, violating P-invariance and, possibly,
T-invariance, can be accounted in the scheme of coupled channels
as perturbation. We separate from the operator of neutron-nucleon
interaction a small component $\delta\hat{U}$
\beq{4.1}
\hat{U} \longrightarrow \hat{U}+\delta \hat{U}.
\eeq
We assume that the operator $\hat{U}$ on the right side of
Eq.(\ref{4.1}) is T-invariant.

A perturbation $\delta\hat{U}$ results in small change of radial
functions and S-matrix elements
\beq{4.2}
F^J_{n_0n}(r)\longrightarrow F^J_{n_0n}(r)+
\delta F^J_{n_0n}(r),\quad
S_J(n_0\to n)\longrightarrow S_J(n_0\to n)+
\delta S_J(n_0\to n).
\eeq
Substituting Eqs.(\ref{4.1}), (\ref{4.2}) to Eqs.(\ref{2.17}),
(\ref{2.18}), we get in zero order in $\delta\hat{U}$ a
homogeneous set of equations for functions $F^J_{n_0n}(r)$
\beq{4.3}
\frac{d^2F^J_{n_0n}(r)}{dr^2}-
\sum_{n'}q^J_{nn'}(r)F^J_{n_0n'}(r)=0
\eeq
with boundary conditions (\ref{2.18}) and (\ref{2.19}). S-matrix,
determined by T-invariant interaction $\hat{U}$, is
symmetric with respect to the main diagonal
\beq{4.4}
S_J(n_0\to n)=S_J(n\to n_0).
\eeq
We assume that the matrix element of the operator of strong
interaction, $\hat{U}$, between states $|JM;n>$ is a function,
dependent on $r$. So we obtain
\beq{4.5}
q^J_{nn'}(r)=<JM;n|\frac{2m\hat{U}}{\hbar^2}+
\frac{l(l+1)}{r^2}-k^2_n|JM;n'>.
\eeq
This matrix is real and symmetric
\beq{4.6}
q^J_{nn'}(r)=q^{J*}_{n'n}(r)=q^J_{n'n}(r)
\eeq
due to hermiticity, R- and T-invariance of the operator $\hat{U}$
\cite{16a}.

In first order in $\delta\hat{U}$ a non-homogeneous set of
equations for functions $\delta F^J_{n_0n}(r)$ arises
\beq{4.7}
\frac{d^2\delta F^J_{n_0n}(r)}{dr^2}-
\sum_{n'}q^J_{nn'}(r)\delta F^J_{n_0n'}(r)=
Q^J_{n_0n}(r)
\eeq
with boundary conditions
\begin{eqnarray}
&&\rdown{\delta F^J_{n_0n}(r)}=
\frac{1}{2(k_{n_0}k_n)^{1/2}}
\delta S_J(n_0\to n)\,k_nr\,h^{(+)}_l(k_nr),
\label{4.8}\\[\medskipamount]
&&\delta F^J_{n_0n}(0)=0.\label{4.9}
\end{eqnarray}
Sources on the right side of Eqs.(\ref{4.7}) are of the form
\beq{4.10}
Q^J_{n_0n}(r)=r\sum_{n'}
<JM;n|\frac{2m\,\delta \hat{U}}{\hbar^2}|JM;n'>
\frac{F^J_{n_0n'}(r)}{r}.
\eeq

We shall search solutions of the non-homogeneous set of equations
(\ref{4.7}) as series in linearly independent solutions of the
homogeneous set of equations (\ref{4.3}).
We accept for definiteness, that the index $n$ takes on values
of $1,2\ldots N$. It is known, that a set of $N$ differential
equations of second order has $2N$ linearly independent solutions.
For each value $n_0$ the regular solution $F^J_{n_0n}(r)$ of the
set (\ref{4.3}) exists. Besides for each $n_0$ one can construct
an irregular solution, satisfying to condition
\beq{4.11}
\rdown{G^J_{n_0n}(r)}=
\delta_{n_0n}\,k_nr\,h^{(+)}_l(k_nr).
\eeq
As far as $n_0$ also takes on $N$ values ($n_0=1,2\ldots N$), the
functions $F^J_{n_0n}(r)$ and $G^J_{n_0n}(r)$ form the required
fundamental system of $2N$ linearly independent solutions. It is
easy to show, that Wronskian determinant (strokes designate
differentiation with respect to $r$)
\beq{4.12}
W^J(r)=\left|
\begin{array}{ccccccc}
F^J_{11}(r)        & F^J_{21}(r)        & \ldots &
F^J_{N1}(r)        & G^J_{11}(r)        & \ldots &
G^J_{N1}(r)\\
\vdots             & \vdots             & \ddots &
\vdots             & \vdots             & \ddots &
\vdots     \\
F^J_{1N}(r)        & F^J_{2N}(r)        & \ldots &
F^J_{NN}(r)        & G^J_{1N}(r)        & \ldots &
G^J_{NN}(r)\\
F^{J^{\pr}}_{11}(r)& F^{J^{\pr}}_{21}(r)& \ldots &
F^{J^{\pr}}_{N1}(r)& G^{J^{\pr}}_{11}(r)& \ldots &
G^{J^{\pr}}_{N1}(r)\\
\vdots             & \vdots             & \ddots &
\vdots             & \vdots             & \ddots &
\vdots     \\
F^{J^{\pr}}_{1N}(r)& F^{J^{\pr}}_{2N}(r)& \ldots &
F^{J^{\pr}}_{NN}(r)& G^{J^{\pr}}_{1N}(r)& \ldots &
G^{J^{\pr}}_{NN}(r)
\end{array}
\right|,
\eeq
does not depend on $r$ and equals $i^N$.

We write now the solutions of the non-homogeneous set of equations
(\ref{4.7}) in the form
\beq{4.13}
\delta F^J_{n_0n}(r)=
\sum_{n'}\Bigl(\int\limits_0^r
\beta^J_{n_0n'}(r')dr'+
b^J_{n_0n'}\Bigr)F^J_{n'n}(r)+
\sum_{n'}\Bigl(\int\limits_0^r
\gamma^J_{n_0n'}(r')dr'+
c^J_{n_0n'}\Bigr)G^J_{n'n}(r).
\eeq
Substituting these functions to the set (\ref{4.7}), we obtain
$2N$ algebraic equations for $2N$ functions $\beta^J_{n_0n}(r)$
and $\gamma^J_{n_0n}(r)$ for each value of $n_0$
\beq{4.14}
\left\{
\begin{array}{rcl}
\sum\limits_{n'}\beta^J_{n_0n'}(r)F^J_{n'n}(r)+
\sum\limits_{n'}\gamma^J_{n_0n'}(r)G^J_{n'n}(r)&=&0,
\\[\bigskipamount]
\sum\limits_{n'}\beta^J_{n_0n'}(r)F^{J^{\pr}}_{n'n}(r)+
\sum\limits_{n'}
\gamma^J_{n_0n'}(r)G^{J^{\pr}}_{n'n}(r)&=&Q^J_{n_0n}(r).
\end{array}
\right.
\eeq
This set is tractable because of $W^J=i^N$ is not equal to zero.

Constants $b^J_{n_0n}$ and $c^J_{n_0n}$ can be found from the
boundary conditions. Indeed, owing to regularity of the functions
$\delta F^J_{n_0n}(r)$ in the origin we have
\beq{4.15}
c^J_{n_0n}=0.
\eeq
On the other hand, as far as the functions $\delta F^J_{n_0n}(r)$
(\ref{4.8}) outside the interaction region contain only
diverging waves, coefficients at functions $F^J_{n'n}(r)$ should
vanish at $r>R$. This gives
\beq{4.16}
b^J_{n_0n}=-\int\limits_0^R \beta^J_{n_0n}(r)dr.
\eeq
We notice, that an upper limit in this integral may take any value
greater than $R$, as at $r>R$ we have $Q^J_{n_0n}(r)=0$ and,
correspondingly, $\beta^J_{n_0n}(r)=\gamma^J_{n_0n}(r)=0$.

So we get for the functions $\delta F^J_{n_0n}(r)$ outside the
interaction region
\beq{4.17}
\rdown{\delta F^J_{n_0n}(r)}=
\Bigl( \int\limits_0^{\infty}
\gamma^J_{n_0n}(r)dr \Bigr)
\,k_nr\,h^{(+)}_l(k_nr).
\eeq
Comparing this result with Eq.(\ref{4.8}), we find
the first order corrections to the S-matrix
\beq{4.18}
\delta S_J(n_0\to n)=2(k_{n_0}k_n)^{1/2}
\int\limits_0^{\infty} \gamma^J_{n_0n}(r)dr.
\eeq

We are interesting now in the functions $\gamma^J_{n_0n}(r)$.
They can be written in the form
\beq{4.19}
\gamma^J_{n_0n}(r)=\frac{1}{W^J}
\sum_{n'}Q^J_{n_0n'}(r)A^J_{nn'}(r).
\eeq
Here $A^J_{nn'}(r)$ is an algebraic adjoint of an element
$G^{J^{\pr}}_{nn'}(r)$ of the Wronskian determinant (\ref{4.12}).
It is easy to check, that the algebraic adjoints $A^J_{nn'}(r)$
satisfy a homogeneous set of equations
\beq{4.20}
\frac{d^2A^J_{nn'}(r)}{dr^2}-
\sum_{n''}q^J_{n''n'}(r)A^J_{nn''}(r)=0.
\eeq
As far as the matrix $q^J_{n''n'}(r)$ is symmetric, the functions
$A^J_{nn'}(r)$ are solutions of Eqs.(\ref{4.3}). On the other
hand, the adjoints $A^J_{nn'}(r)$ outside the interaction region
take the form
\begin{eqnarray}
\rdown{A^J_{nn}(r)}&=&i^{N-1}\frac{1}{2k_n}
\left(\,k_nr\,h^{(-)}_l(k_nr)+
S_J(n\to n)\,k_nr\,h^{(+)}_l(k_nr)\right),
\label{4.21}\\[\medskipamount]
\rdown{A^J_{nn'}(r)}&=&i^{N-1}\frac{1}{2(k_nk_{n'})^{1/2}}
S_J(n'\to n)\,k_{n'}r\,h^{(+)}_{l'}(k_{n'}r),\quad
n\ne n'. \label{4.22}
\end{eqnarray}
In view of the symmetry of the S-matrix (\ref{4.4}), these
functions $A^J_{nn'}(r)$ are proportional to the regular solutions
$F^J_{nn'}(r)$ (\ref{2.18}) at $r>R$. Thus, we conclude, that for
any value of $r$ a relationship holds
\beq{4.23}
A^J_{nn'}(r)=i^{N-1}F^J_{nn'}(r),
\eeq
because of the functions $A^J_{nn'}(r)$ and $F^J_{nn'}(r)$ are
solutions of the same set of equations (\ref{4.3}).

Substituting explicit expressions for the sources $Q^J_{n_0n}(r)$
(\ref{4.10}) and the algebraic adjoints $A^J_{nn'}(r)$
(\ref{4.23}) to Eq.(\ref{4.19}), we obtain the functions
$\gamma^J_{n_0n}(r)$. Thus the first order corrections to the
S-matrix are of the form
\begin{eqnarray}
&&\delta S_J(n_0\to n)=-2i(k_{n_0}k_n)^{1/2}
\times{}\nonumber\\[\medskipamount]
&&{}\times\sum_{n'n''}\int\limits_{0}^{\infty}r^2dr
\left(\frac{F^J_{nn'}(r)}{r}
<JM;n'|\frac{2m\delta\hat{U}}{\hbar^2}|JM;n''>
\frac{F^J_{n_0n''}(r)}{r}\right).\label{4.24}
\end{eqnarray}

If the interaction $\delta\hat{U}$ is T-invariant, the matrix
element on the right side of Eq.(\ref{4.24}) is symmetric,
therefore we have
\beq{4.25}
\delta S_J(n_0\to n)=\delta S_J(n\to n_0).
\eeq
While if $\delta\hat{U}$ anticommutates with the operator of time
inversion, $\hat{T}$, then the appropriate matrix element is
antisymmetric, so
\beq{4.26}
\delta S_J(n_0\to n)=-\delta S_J(n\to n_0).
\eeq
In a general case, when the interaction $\delta\hat{U}$ includes
both T-invariant and T-noninvariant components, the S-matrix
correction is a sum of symmetric and antisymmetric matrixes.

Taking into account the definition (\ref{2.4}), we present the
matrix element of the operator $\delta\hat{U}$ in the form
\begin{eqnarray}
&&<JM;n|\delta\hat{U}|JM;n'>=
\sum_{\nu\mu\nu'\mu'}C^{JM}_{j\nu I\mu}C^{JM}_{j'\nu' I'\mu'}
\times{}\nonumber\\[\medskipamount]
&&\phantom{<JM;n|\delta\hat{U}|JM;n'>}\times
<j\nu;l|\left(\int\nolimits d\tau \psi^*_{\al}(\tau)
\delta\hat{U}\psi_{\al'}(\tau)\right)|j'\nu';l'>.\label{4.27}
\end{eqnarray}
Let us consider in greater detail a simplest variant, when
integrals over internal variables $\tau$ are single-particle
operators. If the interaction $\delta\hat{U}$ violates
P-invariance only, then
\beq{4.28}
\int d\tau \psi^*_{\al}(\tau)
\delta\hat{U}_P\psi_{\al'}(\tau)=
\delta_{II'}\delta_{\mu\mu'}\delta_{\pi\pi'}
\frac{R}{2\hbar}
\left(U^{(ii')}_P(r)\hat{\sig}\hat{\bf p}+
\hat{\sig}\hat{\bf p}U^{(ii')}_P(r)\right),
\eeq
where $\hat{\sig}=2\hat{\bf s}$, $\hat{\bf s}$ is an operator of
a neutron spin, and $\hat{\bf p}=-i\hbar\partial /\partial{\bf r}$
is a momentum operator. Function $U^{(ii')}_P(r)$ has dimensions
of energy owing to a factor $R/\hbar$. Similarly,
a single-particle operator, violating P- and T-invariance, takes
the form
\beq{4.29}
\int d\tau \psi^*_{\al}(\tau)
\delta\hat{U}_{PT}\psi_{\al'}(\tau)=
\delta_{II'}\delta_{\mu\mu'}\delta_{\pi\pi'}
\frac{1}{2R}
\left(U^{(ii')}_{PT}(r)\hat{\sig}{\bf r}+
\hat{\sig}{\bf r}U^{(ii')}_{PT}(r)\right).
\eeq
Functions $U^{(ii')}_P(r)$ and $U^{(ii')}_{PT}(r)$ are real and
symmetric.

Matrix element of the single-particle P-noninvariant operator
(\ref{4.28}) is given by formula
\beq{4.30}
<j\nu;l|\hat{\sig}\hat{\bf p}|j'\nu';l'>f(r)=
\delta_{jj'}\delta_{\nu\nu'}\sqrt{3}i^{l-l'}
U(jl'\oh 1,\oh l)(l\Vert f(r)\Vert l'),
\eeq
where $U(abcd,ef)=((2e+1)(2f+1))^{1/2}W(abcd,ef)$ is a normalized
Racah function \cite{22a}, and $(l\Vert f(r)\Vert l')$ is
a reduced matrix element. We have for s- and p-waves
\begin{eqnarray}
&&(0\Vert f(r)\Vert 0)=(1\Vert f(r)\Vert 1)=0,
\nonumber\\[\medskipamount]
&&(0\Vert f(r)\Vert 1)=
i\hbar\left(\frac{df(r)}{dr}+2\frac{f(r)}{r}\right),\quad
(1\Vert f(r)\Vert 0)=
-\frac{i\hbar}{\sqrt{3}}\frac{df(r)}{dr}. \label{4.31}
\end{eqnarray}
Similarly for P-,T-noninvariant interaction we obtain
\beq{4.32}
<j\nu;l|\hat{\sig}{\bf r}|j'\nu';l'>=
\delta_{jj'}\delta_{\nu\nu'}\sqrt{3}i^{l-l'}
U(jl'\oh 1,\oh l)\left(\frac{2l'+1}{2l+1}\right)^{1/2}
C^{l0}_{10l'0}r.
\eeq

We have already assumed, that the weak interaction,
$\delta\hat{U}$, mixes only the states of a target with the same
spins, $I$, projections, $\mu$, and parities, $\pi$ (see
Eqs.(\ref{4.28}) and (\ref{4.29})). We consider now a case, when
the strong interaction $\hat{U}$ is also diagonal on spins, $I$,
projections, $\mu$, and parities, $\pi$, of the inner target
states, as well as on orbital, $l$, and total, $j$, neutron
angular momenta. This is indeed the case in the model of
neutron-nucleus interaction studied below. So we
write the radial functions in the form
\beq{4.33}
\frac{F^J_{n_0n}(r)}{r}=
\delta_{I_0I}\delta_{\mu_0\mu}\delta_{\pi_0\pi}
\delta_{l_0l}\delta_{j_0j}\psi^{J(i)}_{lj}(r).
\eeq

Corrections to S-matrix, describing elastic transitions from s- to
p-wave and vice versa, caused by the single-particle
P-noninvariant interaction (\ref{4.28}), have then the form
\begin{eqnarray}
&&S^P_J(0\oh\to 1\oh)=S^P_J(1\oh\to 0\oh)=
ik_0R\frac{2m}{\hbar^2}\sum_{ii'}
\int\limits_{0}^{\infty}r^2dr
\Bigl(\psi^{J(i)}_{0\oh}U^{(ii')}_P
\frac{d\psi^{J(i')}_{1\oh}}{dr}-{}
\nonumber\\[\medskipamount]
&&\phantom{S^P_J(0\oh\to 1\oh)=S^P_J(1\oh}
-\frac{\psi^{J(i)}_{0\oh}}{dr}
U^{(ii')}_P\psi^{J(i')}_{1\oh}+
2\psi^{J(i)}_{0\oh}\frac{U^{(ii')}_P}{r}
\psi^{J(i')}_{1\oh}\Bigr). \label{4.34}
\end{eqnarray}
Similarly for corrections, which are due to the P-,T-invariant
interaction (\ref{4.29}), we find
\beq{4.35}
S^{PT}_J(0\oh\to 1\oh)=-S^{PT}_J(1\oh\to 0\oh)=
k_0R\frac{2m}{\hbar^2}\sum_{ii'}
\int\limits_{0}^{\infty}r^2dr
\psi^{J(i)}_{0\oh}\frac{2rU^{(ii')}_{PT}}{R^2}
\psi^{J(i')}_{1\oh}.
\eeq
Thus, calculating in the framework of the model the radial
functions $\psi^{J(i)}_{lj}(r)$, we get the asymmetries of a
total cross section and the angles of spin rotation
(\ref{3.12})-(\ref{3.15}).

The simplest model is, certainly, the model of potential
neutron-nucleus interaction. Earlier P-noninvariant effects,
caused by single-particle potential of the type (\ref{4.28}), have
been studied in Refs.\cite{23a,10b}. In Ref.\cite{23a} a case of
a spherical square well was investigated, while in
Ref.\cite{10b} an optical potential of a Woods-Saxon type was
considered. To account for the elastic channel only one should
remove in the formulas (\ref{4.34}), (\ref{4.35}) the sums over
excited states $i$ and $i'$ of a target. Thus substituting the
S-matrix elements,
$S^P_J(0\oh\to 1\oh)=S^P_J(1\oh\to 0\oh)$,
to Eq.(\ref{3.12}), we reproduce the expression for P-noninvariant
asymmetry of a total cross section, used in Ref.\cite{10b}.

\section{Model for neutron resonance}
\label{s5}

\subsection {Exact expressions}
\label{s51}

Let us consider a simple model for a resonance, resulted from
a coupling of an incident neutron with one of excited levels of a
target. In an elastic channel a neutron interacts with a nucleus,
being in a ground state with a spin $I$, parity $\pi$ and energy
$\ve_0$. An energy $\ve=E-\ve_0$ of a relative motion is expressed
by a momentum of an incident neutron, $\hbar k_0$, in the ordinary
way $\ve=(\hbar k_0)^2/2m$ (see Eq.(\ref{2.11})). We assume for
simplicity, as was already stated in the previous section, that in
the process of target excitation spin, $I$, projection, $\mu$, and
parity, $\pi$, do not change (see Eq.(\ref{4.33})). It means, that
the matrix elements of the operator of P- and T-invariant strong
interaction, $\hat{U}$, entering in Eq.(\ref{4.3}), is diagonal on
$I$, $\pi$ and, therefore, $j$ and $l$
\beq{5.1}
<JM;n|\hat{U}|JM;n'>=\delta_{II'}\delta_{\pi\pi'}
\delta_{jj'}\delta_{ll'} U^{(ii')}_{ljJ}(r).
\eeq
Also used is an independence of the matrix element on the quantum
number $M$ due to R-invariance of the interaction $\hat{U}$. Thus,
orbital, $l$, and total, $j$, neutron angular momenta do not
change in the model during the interaction with a nucleus (recall,
we are dealing now only with the strong interaction). Potentials
$U^{(ii')}_{ljJ}(r)$ are spherically symmetric, as they result
from an integration over all internal variables of a target and
spin-angular variables of an incident neutron. According to
Eq.(\ref{4.6}) these potentials are real and symmetric on indexes
$i$ and $i'$.

We assume also, that there exists only one inelastic channel in
each partial wave with quantum numbers $l$, $j$ and $J$, related
with an excitation of a target level with an energy $\ve_{1ljJ}$.
Thus we obtain in each partial wave a scattering problem with two
coupled channels --- elastic and inelastic. For simplicity we
shall use only an index $l$ to label partial wave, omitting $j$
and $J$. Let $\ve_l=\ve_{1ljJ}-\ve_0$ is an energy of target
excitation in the partial wave $l$.

We introduce simplified designations for radial functions:
$F^{(0)}_l(r)=F^J_{n_0n_0}(r)$ in the elastic channel and
$F^{(1)}_l(r)=F^J_{n_0n_1}(r)$ in the inelastic one. According to
Eqs.(\ref{4.3}) and (\ref{4.5}) these functions satisfy the
equations
\beq{5.2}
\frac{d^2F^{(n)}_l}{dr^2}-
\frac{l(l+1)}{r^2}F^{(n)}_l-
\sum_{n'} \frac{2mU^{(nn')}_l(r)}{\hbar^2}F^{(n')}_l+
k^2_{nl}F^{(n)}_l=0,
\eeq
where $n,n'=0,1$. We have for wave numbers $k_0$ and $k_{1l}$ in
the elastic and inelastic channels (see Eqs.(\ref{2.11}) and
(\ref{2.12}))
\beq{5.3}
k_0=\frac{1}{R} \left(\frac{\ve}{\om}\right)^{1/2}, \qquad
k_{1l}=\frac{1}{R} \left(\frac{\ve-\ve_l}{\om}\right)^{1/2}.
\eeq
A quantity
\beq{5.4}
\om=\frac{\hbar^2}{2mR^2}
\eeq
is on the scale of one-particle excitation energy for potential of
characteristic radius $R$. We rewrite the boundary conditions
(\ref{2.18}) for the radial functions in the form
\beq{5.5}
\rdown{F^{(n)}_l(r)}=\frac{1}{2(k_0k_{nl})^{1/2}}
\Bigl(\delta_{n0}\,k_{nl}r\,h^{(-)}_l(k_{nl}r)+
S^{(n)}(l \to l)\,k_{nl}r\,h^{(+)}_l(k_{nl}r)\Bigr),
\eeq
where the S-matrix elements $S^{(0)}(l\to l)$ describe elastic
scattering, while \mbox{$S^{(1)}(l\to l)$} correspond to
transitions from the elastic channel to the inelastic one.

We take all the potentials to be square wells of the same radius
$R$
\begin{eqnarray}
U^{(nn)}_l(r)&=&\left\{
\begin{array}{rr}
         0, & r>R;\\
-U^{(n)}_l, & r<R;\\
\end{array}
\right.
\label{5.6} \\[\medskipamount]
U^{(01)}_l(r)=U^{(10)}_l(r)&=&\left\{
\begin{array}{rr}
   0, & r>R;\\
-W_l, & r<R.\\
\end{array}
\right. \label{5.7}
\end{eqnarray}
Then the equations (\ref{5.2}) can be solved analytically. Regular
solutions in the region $r< R$ are of the form
\beq{5.8}
\left\{
\begin{array}{rcl}
F^{(0)}_l(r)&=&\left(\frac{\tst R}{\tst k_0}\right)^{1/2}
\Bigl( A_l\,\kp_lr\,j_l(\kp_lr)+
      A'_l\,\kp'_lr\,j_l(\kp'_lr)\Bigr) ,
\\[\bigskipamount]
F^{(1)}_l(r)&=&\left(\frac{\tst R}{\tst k_0}\right)^{1/2}
\Bigl( B_l\,\kp_lr\,j_l(\kp_lr)+
      B'_l\,\kp'_lr\,j_l(\kp'_lr)\Bigr) .
\end{array}
\right.
\eeq
Recall that the spherical Bessel functions, $j_l(x)$, satisfy
the equation
\beq{5.9}
\frac{d^2}{dx^2}(xj_l(x))-
\left(\frac{l(l+1)}{x^2}-1\right)xj_l(x)=0.
\eeq
A factor $(R/k_0)^{1/2}$ is separated in the formulas (\ref{5.8})
to simplify final expressions.

Substituting solutions (\ref{5.8}) in Eqs.(\ref{5.2}) and equating
factors at Bessel functions of the same argument, we get
\beq{5.10}
\left\{
\begin{array}{rcl}
-A_l\kp^2_l+A_l\frac{\tst 2m(U^{(0)}_l+\ve)}{\tst \hbar^2}+
B_l\frac{\tst 2mW_l}{\tst \hbar^2}&=&0,
\\[\bigskipamount]
-B_l\kp^2_l+B_l\frac{\tst 2m(U^{(1)}_l+\ve-\ve_l)}
{\tst \hbar^2}+A_l\frac{\tst 2mW_l}{\tst \hbar^2}&=&0,
\end{array}
\right.
\eeq
and precisely the same set of equations for $A'_l$ and $B'_l$. The
set (\ref{5.10}) is tractable if its determinant equals zero. This
gives the quadratic equation for a quantity $\kp^2_l$, two roots
of which are $\kp^2_l$ and $\kp'^2_l$. We find for $\kp_l$ and
$\kp'_l$
\beq{5.11}
\kp_l=
\left(\frac{2m}{\hbar^2}(U^{(0)}_l+\ve+\Delta_l)\right)^{1/2}=
\frac{1}{R}\left(\frac{U^{(0)}_l+\ve+\Delta_l}{\om}\right)^{1/2},
\eeq
\beq{5.12}
\kp'_l=
\left(\frac{2m}{\hbar^2}(U^{(1)}_l+\ve-\ve_l-
\Delta_l)\right)^{1/2}=
\frac{1}{R}\left(\frac{U^{(1)}_l+\ve-\ve_l-
\Delta_l}{\om}\right)^{1/2},
\eeq
where
\beq{5.13}
\Delta_l=\left(\left(\frac{U^{(0)}_l-U^{(1)}_l+\ve_l}{2}\right)^2+
W^2_l\right)^{1/2}-
\left(\frac{U^{(0)}_l-U^{(1)}_l+\ve_l}{2}\right).
\eeq
We assume that
\beq{5.14}
U^{(0)}_l-U^{(1)}_l+\ve_l>0,
\eeq
therefore in the absence of channel coupling $W_l \to 0$ we have
$\Delta_l \to 0$, so $\kp_l$ and $\kp'_l$ become equal to the wave
numbers in the elastic and inelastic channels, respectively, in
the region $r<R$. Factors $B_l$ and $B'_l$ take the form
\beq{5.15}
B_l=\lal A_l, \qquad B'_l=-\frac{1}{\lal}A'_l,
\eeq
where
\beq{5.16}
\lal=\frac{\Delta_l}{W_l}.
\eeq

Functions (\ref{5.5}) and (\ref{5.8}), and their first derivatives
should be joined at the point $r=R$. This gives four equations for
four unknown quantities $A_l$, $A'_l$, $S^{(0)}(l\to l)$ and
$S^{(1)}(l\to l)$. As far as a relation
\beq{5.17}
\frac{d}{dx}(x^{l+1}f_l(x))=x^{l+1}f_{l-1}(x),
\eeq
is valid for spherical Bessel and Hankel functions, it is
convenient to join the first derivatives of the radial functions,
multiplied by $r^l$.

We get for joining conditions in the inelastic channel
\beq{5.18}
\left\{
\begin{array}{l}
B_l\,\kp_lR\,j_l(\kp_lR)+B'_l\,\kp'_lR\,j_l(\kp'_lR)=
\frac{\tst 1}{\tst 2(k_{1l}R)^{1/2}}
S^{(1)}(l \to l)\,k_{1l}R\,h^{(+)}_l(k_{1l}R),
\\[\bigskipamount]
B_l(\kp_lR)^2j_{l-1}(\kp_lR)+B'_l(\kp'_lR)^2j_{l-1}(\kp'_lR)={}
\\[\bigskipamount]
\phantom{B_l(\kp_lR)^2j_{l-1}(\kp_lR)+B'_l(\kp'_lR)}=
\frac{\tst 1}{\tst 2(k_{1l}R)^{1/2}}
S^{(1)}(l \to l)(k_{1l}R)^2h^{(+)}_{l-1}(k_{1l}R).
\end{array}
\right.
\eeq
Consider a case, when the inelastic channel is closed, that is,
$\ve <\ve_l$, and the channel coupling is absent, so $B_l=0$. Then
Eqs.(\ref{5.18}) specify an energy $\ve-\ve_l < 0$ of a bound
state of a neutron in the square potential of radius $R$ and depth
$U^{(1)}_l$. This energy can be found by equating a determinant
\beq{5.19}
D'_l(\ve)=\,\kp'_lR\,j_l(\kp'_lR)
(k_{1l}R)^2h^{(+)}_{l-1}(k_{1l}R)-
(\kp'_lR)^2j_{l-1}(\kp'_lR)
\,k_{1l}R\,h^{(+)}_l(k_{1l}R).
\eeq
to zero. We introduce by analogy a function
\beq{5.20}
D_l(\ve)=\,\kp_lR\,j_l(\kp_lR)
(k_{1l}R)^2h^{(+)}_{l-1}(k_{1l}R)-
(\kp_lR)^2j_{l-1}(\kp_lR)
\,k_{1l}R\,h^{(+)}_l(k_{1l}R),
\eeq
and return to the analysis of a general case. Thus eliminating
$S^{(1)}(l \to l)$ from Eqs.(\ref{5.18}), we get a relation
between the factors $B_l$ and $B'_l$
\beq{5.21}
B_lD_l(\ve)+B'_lD'_l(\ve)=0.
\eeq

We turn now to joining conditions in the elastic channel
\beq{5.22}
\left\{
\begin{array}{l}
A_l\,\kp_lR\,j_l(\kp_lR)+A'_l\,\kp'_lR\,j_l(\kp'_lR)={}
\\[\bigskipamount]
\phantom{A_l(\kp_lR)}=
\frac{\tst 1}{\tst 2(k_0R)^{1/2}}
\Bigl( \,k_0R\,h^{(-)}_l(k_0R)+
S^{(0)}(l \to l)\,k_0R\,h^{(+)}_l(k_0R) \Bigr) ,
\\[\bigskipamount]
A_l(\kp_lR)^2j_{l-1}(\kp_lR)+A'_l(\kp'_lR)^2j_{l-1}(\kp'_lR)={}
\\[\bigskipamount]
\phantom{A_l(\kp_lR)}=
\frac{\tst 1}{\tst 2(k_0R)^{1/2}}
\Bigl( (k_0R)^2h^{(-)}_{l-1}(k_0R)+
S^{(0)}(l \to l)(k_0R)^2h^{(+)}_{l-1}(k_0R) \Bigr) .
\end{array}
\right.
\eeq
Eliminating $S^{(0)}(l \to l)$ and taking into account, that
\beq{5.23}
h^{(-)}_l(x)h^{(+)}_{l-1}(x)-
h^{(-)}_{l-1}(x)h^{(+)}_l(x)=\frac{2i}{x^2},
\eeq
we obtain a relation between the factors $A_l$ and $A'_l$
\beq{5.24}
\begin{array}{l}
A_l\Bigl( \,\kp_lR\,j_l(\kp_lR)(k_0R)^2h^{(+)}_{l-1}(k_0R)-
(\kp_lR)^2j_{l-1}(\kp_lR)\,k_0R\,h^{(+)}_l(k_0R) \Bigr) + {}
\\[\bigskipamount]
{}+A'_l\Bigl( \,\kp'_lR\,j_l(\kp'_lR)(k_0R)^2h^{(+)}_{l-1}(k_0R)-
(\kp'_lR)^2j_{l-1}(\kp'_lR)\,k_0R\,h^{(+)}_l(k_0R) \Bigr) ={}
\\[\bigskipamount]
\phantom{{}+A'_l\Bigl( (\kp'_lR)
j_l(\kp'_lR)(k_0R)^2h^{(+)}_{l-1}(k_0R)-(}=i(k_0R)^{1/2}.
\end{array}
\eeq
This equation along with (\ref{5.15}) and (\ref{5.21}) enables
us to find the factors $A_l$ and $A'_l$ (and, certainly, $B_l$ and
$B'_l$)
\beq{5.25}
A_l=i(k_0R)^{1/2}\frac{D'_l(\ve)}{Z_l(\ve)}, \qquad
A'_l=i(k_0R)^{1/2}\frac{\lal^2 D_l(\ve)}{Z_l(\ve)},
\eeq
where
\beq{5.26}
\begin{array}{l}
Z_l(\ve)=
(k_0R)^2h^{(+)}_{l-1}(k_0R)
\Bigl( D'_l(\ve)\,\kp_lR\,j_l(\kp_lR)+
\lal^2 D_l(\ve)\,\kp'_lR\,j_l(\kp'_lR) \Bigr) - {}
\\[\bigskipamount]
{}-\,k_0R\,h^{(+)}_l(k_0R)
\Bigl( D'_l(\ve)(\kp_lR)^2j_{l-1}(\kp_lR)+
\lal^2 D_l(\ve)(\kp'_lR)^2j_{l-1}(\kp'_lR) \Bigr) .
\end{array}
\eeq
Formulas (\ref{5.15}) and (\ref{5.25}) specify the
energy dependent factors $A_l$, $A'_l$, $B_l$ and $B'_l$, and,
therefore, radial wave functions (\ref{5.8}).

We find now a position of a resonance in the partial wave $l$. A
cross section for elastic scattering is determined by an element
of S-matrix, $S^{(0)}(l \to l)$, or a phase shift, $\delta_l$.
These quantities express in terms of a logarithmic derivative of
the elastic-channel radial function in the ordinary way \cite{16a}
\beq{5.27}
S^{(0)}(l \to l)=\exp(2i\delta_l)=
\exp(2i\delta^c_l)
\frac{\Phi_l(\ve)+is_l}{\Phi_l(\ve)-is_l},
\eeq
where
\beq{5.28}
\Phi_l(\ve)=\tilde{\Phi}_l(\ve)-d_l, \qquad
\tilde{\Phi}_l(\ve)=R\mathop{\frac{dF^{(0)}_l/dr}{F^{(0)}_l}}
\nolimits_{|R}.
\eeq
This results from the joining conditions (\ref{5.22}).
Phase shifts, $\delta^c_l$, for scattering on an opaque sphere of
radius $R$ are given by equation
\beq{5.29}
\exp(2i\delta^c_l)=-\frac{h^{(-)}_l(k_0R)}{h^{(+)}_l(k_0R)},
\eeq
while factors of shift, $d_l(\ve)$, and penetrability, $s_l(\ve)$,
are real and imaginary parts of expression
\beq{5.30}
d_l+is_l=\mathop{\frac{(xh^{(+)}_l(x))^{\pr}}{h^{(+)}_l(x)}}
\nolimits_{|x=k_0R}=
-l+\frac{(k_0R)h^{(+)}_{l-1}(k_0R)}{h^{(+)}_l(k_0R)}.
\eeq
At low energies ($k_0R \ll 1$ ) the following asymptotic formulas
are
\beq{5.31}
\delta^c_l \simeq -\frac{(k_0R)^{2l+1}}{(2l-1)!!(2l+1)!!}, \qquad
d_l \simeq -l, \qquad
s_l \simeq \frac{(k_0R)^{2l+1}}{((2l-1)!!)^2}.
\eeq

A resonance occurs at an energy $E_l$, such that $\Phi_l(E_l)=0$.
If we restrict to the first term of the Taylor expansion
\beq{5.32}
\Phi_l(\ve)=-\frac{1}{\gamma_l}(\ve-E_l),
\eeq
we find the S-matrix element in the Breit-Wigner form
\beq{5.33}
S^{(0)}(l \to l)=\exp(2i\delta^c_l) \Bigl(
1-\frac{i\Gamma_l}
{\ve-E_l+i\frac{\tst \Gamma_l}{\tst \mathstrut 2}} \Bigr).
\eeq
A width of a resonance is specified by expression
\beq{5.34}
\Gamma_l=2s_l\gamma_l,
\eeq
where $\gamma_l$ is a reduced width.

In the model that we consider the logarithmic derivative of the
elastic-channel function $F^{(0)}_l(r)$ at the point $r=R$ takes
the form
\beq{5.35}
\Phi_l(\ve)=
\frac{A_l(\kp_lR)^2j_{l-1}(\kp_lR)+
A'_l(\kp'_lR)^2j_{l-1}(\kp'_lR)}
{A_l\,\kp_lR\,j_l(\kp_lR)+
A'_l\,\kp'_lR\,j_l(\kp'_lR)}-l-d_l.
\eeq
Substituting to this formula the energy dependent factors $A_l$
and $A'_l$, we obtain the explicit expression for the logarithmic
derivative $\Phi_l(\ve)$ and, therefore, for the phase shift
$\delta_l$ and the partial cross section for elastic scattering.

\subsection {Approximate expressions}
\label{s52}

In this section we study in greater details an energy dependence
of the factors $A_l$, $A'_l$, $B_l$ and $B'_l$ near a narrow
resonance. Such resonance occurs, if a coupling between channels
is weak, that is, if $|W_l|\ll \ve_l$. Provided
$U^{(0)}_l\sim U^{(1)}_l$, we have $\Delta_l \simeq W^2_l/\ve_l$,
so the dimensionless parameter $\lal$ (\ref{5.16}) is small
$|\lal|\sim |W_l/\ve_l|\ll 1$. We assume that the energy of
incident neutrons is low ($k_0R \ll 1$), so we use the asymptotic
formulas (\ref{5.31}).

Returning to the exact expression (\ref{5.35}) for the logarithmic
derivative of the elastic-channel wave function, we notice, that
$A'_l=0$ in the absence of channel coupling. Then
Eq.(\ref{5.35}) specifies the logarithmic derivative of the radial
function for the elastic scattering of a neutron on a spherical
potential well of radius $R$ and depth $U^{(0)}_l$
\beq{5.36}
\Phi^0_l(\ve)=
\frac{\kp_lR\,j_{l-1}(\kp_lR)}{j_l(\kp_lR)}.
\eeq
Let us show, that the exact expression (\ref{5.35}) for
$\Phi_l(\ve)$ reduces to $\Phi^0_l(\ve)$ (\ref{5.36}) everywhere,
except in a small vicinity of a resonance.

Taking into account Eqs.(\ref{5.25}), we rewrite the expression
(\ref{5.35}) in the form
\beq{5.37}
\Phi_l(\ve)=
\frac{D'_l(\ve)(\kp_lR)^2j_{l-1}(\kp_lR)+
\lal^2 D_l(\ve)(\kp'_lR)^2j_{l-1}(\kp'_lR)}
{D'_l(\ve)\,\kp_lR\,j_l(\kp_lR)+
\lal^2 D_l(\ve)\,\kp'_lR\,j_l(\kp'_lR)}.
\eeq
We assume that there are no peculiarities in the potential
scattering (i.e., in the quantity (\ref{5.36})) in the considered
region of low energy. Then, obviously, at $\lal^2 \ll 1$
function $\Phi_l(\ve)$ coincides with the accuracy of $\lal^2$
with $\Phi^0_l(\ve)$ everywhere, except in a small vicinity of
an energy $E'_l$, such that $D'_l(E'_l)=0$. We note, that at
$\lal=0$ there exists a bound state of a neutron with an energy
$E'_l-\ve_l$ in the inelastic channel (see text before
Eq.(\ref{5.19})). So, clearly, the scattering cross section
deviates from the potential behavior and, as we shall see, has the
resonant nature just in the vicinity of the energy $E'_l$.

We find now an energy $E_l$ of a resonance. According to equations
(\ref{5.3}), (\ref{5.11}), (\ref{5.12}) and definitions
(\ref{5.19}), (\ref{5.20}) all functions, entering into numerator
and denominator of the fraction (\ref{5.37}), are slow functions
of energy. Namely, they change significantly on the
scale of $\om$ (\ref{5.4}). Therefore the quantity $\Phi_l(\ve)$
(\ref{5.37}) is a slow function of energy everywhere, except in a
small vicinity of the energy $E'_l$. In this vicinity we present
the function $D'_l(\ve)$ in the form
\beq{5.38}
D'_l(\ve)=(\ve-E'_l)
\mathop{\frac{dD'_l(\ve)}{d\ve}}\nolimits_{|E'_l},
\eeq
that is valid, incidentally, on an interval $|\ve-E'_l|\ll\om$.
The derivative, $dD'_l(\ve)/d\ve$, is on the order of
$\sim D'_l(\ve)/\om$. It is easy to show, that the numerator of
the fraction (\ref{5.37}) becomes zero at the energy
\beq{5.39}
E_l=E'_l-\lal^2
\frac{1}
{\mathop{\tst \left(\frac{\tst dD'_l(\ve)}{\tst d\ve}\right)}
\nolimits_{E'_l}}
\mathop{\left(\frac{D_l(\ve)(\kp'_lR)^2j_{l-1}(\kp'_lR)}
{(\kp_lR)^2j_{l-1}(\kp_lR)}\right)}\nolimits_{E_l},
\eeq
and the difference $|E_l-E'_l|$ is small in comparison
with $\om$ owing to $\lal^2 \ll 1$. Therefore
$\mathop{dD'_l(\ve)/d\ve}\nolimits_{|E'_l}\simeq
\mathop{dD'_l(\ve)/d\ve}\nolimits_{|E_l}$. The energy $E_l$
(\ref{5.39}) is the required energy of a resonance in the partial
wave $l$, occurring due to a coupling between the elastic and
inelastic channels.

We study now a behavior of the logarithmic derivative near the
energy $E_l$. Substituting the expansion (\ref{5.38}) to
Eq.(\ref{5.37}) and taking into account the definition
(\ref{5.36}) and result (\ref{5.39}), we obtain
\beq{5.40}
\Phi_l(\ve)=\frac{\ve-E_l}
{\displaystyle \frac{\mathstrut \ve-E_l}{\Phi^0_l}-\gamma_l},
\eeq
where $\Phi^0_l=\Phi^0_l(E_l)$ and
\beq{5.41}
\gamma_l=\lal^2 \om \mathop{\left(
\frac{2\om D^2_l(\ve)(\kp'_lR)^2}
{\tst (U^{(1)}_l-\Delta_l)
\left( (\kp_lR)^2j_{l-1}(\kp_lR)\right)^2
(k_{1l}R)^2 h^{(+)}_{l-1}(k_{1l}R)h^{(+)}_{l+1}(k_{1l}R)}
\right)}\nolimits_{E_l}.
\eeq
We see, that close to $E_l$ the energy dependence of the function
$\Phi_l(\ve)$ is determined by the quantity
$\gamma_l \sim \lal^2 \om \ll \om$. In the small vicinity of $E_l$
\beq{5.42}
|\ve-E_l| \ll \gamma_l |\Phi^0_l|
\eeq
we get the Breit-Wigner formulas (\ref{5.32})-(\ref{5.34}). We
note, however, that the parametrization (\ref{5.40}) holds for
much more wide energy interval $|\ve-E_l|\ll\om$. Thus in the
region, lying beyond the reduced width of a
resonance,
\beq{5.43}
\gamma_l|\Phi^0_l| \ll |\ve-E_l| \ll \om,
\eeq
the logarithmic derivative $\Phi_l(\ve)$ (\ref{5.40}) reduces
simply to the potential value (\ref{5.36}), taken at the point
$E_l$.

Similarly, substituting the expansion (\ref{5.38}) in
Eqs.(\ref{5.25}), (\ref{5.26}) and taking into account
Eqs.(\ref{5.15}) and (\ref{5.41}), we obtain the following
expressions
for the factors $A_l$, $A'_l$, $B_l$ and $B'_l$ on the interval
$|\ve-E_l|\ll\om$
\beq{5.44}
A_l(\ve)=\exp(i\delta^c_l)(s_l)^{1/2}
\frac{(\ve-E_l-\gamma_lx_l)}{[l]}
\mathop{\left(\frac{1}{(\kp_lR)^2j_{l-1}(\kp_lR)}\right)}
\nolimits_{E_l},
\eeq
\beq{5.45}
A'_l(\ve)=\exp(i\delta^c_l)(s_l)^{1/2}
\frac{\gamma_lx_l}{[l]}
\mathop{\left(\frac{1}{(\kp'_lR)^2j_{l-1}(\kp'_lR)}\right)}
\nolimits_{E_l},
\eeq
\beq{5.46}
B_l(\ve)=y_l\left(\frac{\gamma_l}{\om}\right)^{1/2}A_l(\ve),
\qquad B'_l(\ve)=-\frac{1}{y_l}
\left(\frac{\om}{\gamma_l}\right)^{1/2}A'_l(\ve).
\eeq
Here the designations are introduced for quantities
\beq{5.47}
x_l=\mathop{\left(
\frac{(\kp_lR)^2j_{l-1}(\kp_lR)(k_{1l}R)^2
h^{(+)}_{l-1}(k_{1l}R)}{D_l(\ve)}\right)}
\nolimits_{E_l},
\eeq
\beq{5.48}
y_l=\mathop{\left(\frac{x_l(U^{(1)}_l-\Delta_l)
(\kp_lR)^2j_{l-1}(\kp_lR)
h^{(+)}_{l+1}(k_{1l}R)}
{2\om D_l(\ve)(\kp'_lR)^2}\right)}^{1/2}
\nolimits_{E_l},
\eeq
as well as for a resonance denominator
\beq{5.49}
[l]=\ve-E_l+is_l
\left(\gamma_l-\frac{\ve-E_l}{\Phi^0_l}
\right).
\eeq

A factor, containing this denominator, has the following
asymptotic forms
\beq{5.50}
\frac{1}{[l]} \to \left\{
\begin{array}{lcl}
{\displaystyle \frac{1}
{\ve-E_l+i\frac{\tst \Gamma_l}{\tst \mathstrut 2}}},&\mbox{if}&
|\ve-E_l| \ll \gamma_l |\Phi^0_l|;
\\[\bigskipamount]
{\displaystyle \frac{1}{\ve-E_l}
\exp \left( i\frac{s_l}{\Phi^0_l}\right)},
&\mbox{if}&
\gamma_l |\Phi^0_l| \ll |\ve-E_l| \ll \om.
\end{array}
\right.
\eeq
Due to this factor all quantities (\ref{5.44})-(\ref{5.46})
increase resonantly in the small vicinity of $E_l$. Comparing the
coefficients at this factor, however, it is easy to show, that
\beq{5.51}
|B_l| \ll |A_l| \sim |A'_l| \ll |B'_l|, \quad
\mbox{if} \quad
|\ve-E_l| \ll \gamma_l |\Phi^0_l|.
\eeq

We consider now the region (\ref{5.43}), lying outside the
resonance. Clearly, the factors $A_l$ and $B_l$ take the constant
values, while $A'_l$ and $B'_l$ fall off with a spacing from
the resonance as follows $\sim 1/(\ve-E_l)$. Therefore in this
region
\beq{5.52}
|A'_l| \ll |A_l|, \quad \mbox{if} \quad
\gamma_l|\Phi^0_l| \ll |\ve-E_l| \ll \om.
\eeq
The superiority of $B'_l$ over $A_l$ is still persist in the
interval, which goes far beyond the reduced width of the resonance
\beq{5.53}
|A_l| \ll |B'_l|, \quad \mbox{if} \quad
|\ve-E_l| \ll \gamma^{1/2}_l \om^{1/2}
\frac{x_l}{y_l}
\mathop{\left| \frac{(\kp_lR)^2j_{l-1}(\kp_lR)}
{(\kp'_lR)^2j_{l-1}(\kp'_lR)}\right|}\nolimits_{E_l}.
\eeq
On the other hand, the factor $B'_l$ surpasses $B_l$ practically
everywhere over the region of applicability of
Eqs.(\ref{5.44})-(\ref{5.46})
\beq{5.54}
|B_l| \ll |B'_l|, \quad \mbox{if} \quad
|\ve-E_l| \ll \om
\frac{x_l}{y^2_l}
\mathop{\left| \frac{(\kp_lR)^2j_{l-1}(\kp_lR)}
{(\kp'_lR)^2j_{l-1}(\kp'_lR)}\right|}\nolimits_{E_l}.
\eeq

Thus, far away from the resonance
($|\ve-E_l|>\gamma^{1/2}_l\om^{1/2}$) the term, proportional to
the factor $A_l$, dominates in the wave function (\ref{5.8}).
Here, we are dealing with potential elastic scattering. The
probability density to find a neutron inside a target is
proportional to the factor of penetrability of centrifugal barrier
and potential jump at a nuclear boundary \cite{24a}
\beq{5.55}
\mathop{\left( |A_l|^2 \right)}\nolimits_{pot}
\sim s_l.
\eeq
While in the resonance ($\ve \sim E_l$ ) we have for the
probability density in the elastic and inelastic channels
\beq{5.56}
\mathop{\left( |A_l|^2 \right)}\nolimits_{res} \sim
\mathop{\left( |A'_l|^2 \right)}\nolimits_{res} \sim
\frac{1}{s_l},
\eeq
\beq{5.57}
\mathop{\left( |B'_l|^2 \right)}\nolimits_{res} \sim
\frac{\om}{s_l \gamma_l}.
\eeq

An increase of probability density in the resonance can be
qualitatively explained as follows. The characteristic time of
change of a wave packet, made from one-particle functions inside
a nucleus, is $T=\hbar /\om$. An exit of the packet is
retarded by centrifugal barrier and potential jump. Therefore the
exit time is evaluated by $T_l=T/s_l$, so
\beq{5.58}
\mathop{\left( |A_l|^2 \right)}\nolimits_{res} \sim
\mathop{\left( |A'_l|^2 \right)}\nolimits_{res} \sim
\frac{T_l}{T}.
\eeq
As for the inelastic channel, in the model that we consider an
exit of neutrons is retarded by a weak coupling between the
channels. Taking into account the definition of the resonance
width (\ref{5.34}), we can write
\beq{5.59}
\mathop{\left( |B'_l|^2 \right)}\nolimits_{res} \sim
\frac{\om}{\Gamma_l} \sim \frac{\tau_l}{T},
\eeq
where $\tau_l=\hbar/\Gamma_l$ is the usually defined life time of
compound nucleus. Thus we see, that both in the elastic and
inelastic channels an enhancement in probability density
corresponds to an increase of time spent by a neutron inside a
target.

In closing of this section we note, that the quantities $x_l$ and
$y_l$ are real until the inelastic channel is closed
($\ve < \ve_l$), so the phases of the factors $A_l$, $A'_l$, $B_l$
and $B'_l$ coincide. Moreover, these phases are equal to the phase
shift for elastic scattering in the partial wave $l$
\begin{eqnarray}
\delta_l&=&\delta^c_l+\arg \frac{1}{[l]}=
\delta^c_l+\arctg \frac{s_l}{\Phi_l(\ve)} \to
\nonumber\\[\medskipamount]
&\to&\left\{
\begin{array}{lcl}
{\displaystyle \delta^c_l+
\arctg \frac{\Gamma_l(\ve)}{2(E_l-\ve)}},
&\mbox{if}&
|\ve-E_l| \ll \gamma_l |\Phi^0_l|;
\\[\bigskipamount]
{\displaystyle \delta^c_l+\frac{s_l}{\Phi^0_l}},
&\mbox{if}&
\gamma_l |\Phi^0_l| \ll |\ve-E_l| \ll \om.
\end{array}
\right. \label{5.60}
\end{eqnarray}
In a general way, this results from Eqs.(\ref{5.22}). Indeed,
a factor $\exp (i\delta_l)$ can be separated as a common phase
factor on the right sides of these equations.

\section{P- and T-noninvariant mixing of s- and p-wave resonances}
\label{s6}

In the previous section the model for resonance in the arbitrary
partial wave $l$ was constructed. It is easy to apply the results
obtained to describe a close-lying pair of two narrow
s- and p-wave resonances. The needed spherical Bessel and Hankel
functions are of the form
\beq{6.1}
\begin{array}{lll}
{\displaystyle j_{-1}(x)=\frac{\cos x}{x}},&
{\displaystyle j_0(x)=\frac{\sin x}{x}},&
{\displaystyle j_1(x)=\frac{\sin x}{x^2}-\frac{\cos x}{x}},
\\[\bigskipamount]
{\displaystyle h^{(+)}_{-1}(x)=\frac{e^{ix}}{x}},&
{\displaystyle h^{(+)}_0(x)=\frac{-i}{x}e^{ix}},&
{\displaystyle h^{(+)}_1(x)=
-\left(\frac{1}{x}+\frac{i}{x^2}\right)e^{ix}},
\\[\bigskipamount]
&&{\displaystyle h^{(+)}_2(x)=\left(
\frac{i}{x}-\frac{3}{x^2}-\frac{3i}{x^3}\right)e^{ix}}.
\end{array}
\eeq
We assume, that a distance between resonances is much less than a
one-particle energy $\om$. So we use the result (\ref{5.54}) and
neglect the components $\sim B_l$ of the radial functions in the
inelastic channel (see Eq.(\ref{5.8})).

The parity violating interaction of the form (\ref{4.28}) leads to
the elements of S-matrix (\ref{4.34}), corresponding to
transitions from s- to p-wave. Similarly the interaction of the
form (\ref{4.29}), violating the space parity and time reversal
symmetry, results in the corrections (\ref{4.35}) to S-matrix. The
substitution of the radial functions (\ref{5.8}) to the
Eqs.(\ref{4.34}), (\ref{4.35}) gives the elements of S-matrix,
caused by P- and T-violation. Thus, in the model considered
the transmission asymmetry
$\ep_P$ (\ref{3.17}) and the angle of spin rotation $\chi_P$
(\ref{3.13}) are obtainable from the formulas
\beq{6.2}
\frac{\ep_P}{n}=-\frac{4\pi}{k^2}g_J
{\rm Re} S^P_J(0\oh \to 1\oh), \quad
\frac{\chi_P}{n}=\frac{4\pi}{k^2}g_J
{\rm Im} S^P_J(0\oh \to 1\oh).
\eeq
Similarly the quantities $\ep_{PT}$ (\ref{3.17}) and $\chi_{PT}$
(\ref{3.15}) are determined by the following expressions
\beq{6.4}
\frac{\ep_{PT}}{n}=-\frac{4\pi}{k^2}C_{J\oh}
{\rm Im} S^{PT}_J(0\oh \to 1\oh), \quad
\frac{\chi_{PT}}{n}=-\frac{4\pi}{k^2}C_{J\oh}
{\rm Re} S^{PT}_J(0\oh \to 1\oh).
\eeq

It is apparent here, that P- and T-noninvariant effects enhance
near the resonances. Indeed, it was shown in the previous section
that the amplitudes $A_l$, $A'_l$ and $B'_l$ of radial wave
functions (\ref{5.8}) peak at resonance energies (see
Eqs.(\ref{5.44})-(\ref{5.46})). Thus the radial integrals
(\ref{4.34}), (\ref{4.35}), that is, the corrections to S-matrix,
increase too. To get the explicit expressions for these
corrections the form of the radial dependence of potentials
$U^{(ii')}_P(r)$ and $U^{(ii')}_{PT}(r)$ should be chosen.

By analogy with potentials of strong interaction (see
Eqs.(\ref{5.6}), (\ref{5.7})), we take P- and P-,T-noninvariant
potentials to be spherical square wells of radius $R$
\beq{6.6}
U^{(ii')}_{P(PT)}(r)=\left\{
\begin{array}{ll}
0,&r>R;\\[\bigskipamount]
U^{(ii')}_{P(PT)},&r<R.
\end{array}
\right.
\eeq
Now it is convenient to rewrite the Eq.(\ref{4.34}) in
the form
\begin{eqnarray}
&&S^P_J(0\oh \to 1\oh)=ik_0R \frac{2m}{\hbar^2}
\times{}\nonumber\\[\medskipamount]
&&{}\times\sum_{ii'} \int\limits_0^{\infty} dr F^{(i)}_s
\left\{ 2U^{(ii')}_P(r)
\left(\frac{dF^{(i')}_p}{dr}+\frac{F^{(i')}_p}{r}\right)+
\frac{dU^{(ii')}_P(r)}{dr} F^{(i')}_p \right\},\label{6.7}
\end{eqnarray}
as the differentiation of the step function $U^{(ii')}_P(r)$ gives
the easily integrated \mbox{$\delta$-function}. At the same time
replacing the functions $\psi^{(i)}_l$ by $F^{(i)}_l$ we get for
the amplitude of P-,T-noninvariant transition (\ref{4.35})
\beq{6.8}
S^{PT}_J(0\oh \to 1\oh)=k_0R \frac{2m}{\hbar^2}
\sum_{ii'} \int\limits_0^{\infty} dr F^{(i)}_s
\frac{2rU^{(ii')}_{PT}(r)}{R^2} F^{(i')}_p.
\eeq

Substituting the explicit expressions for radial functions in
these formulas (see Eq.(\ref{5.8})), we can separate contributions
of four types to P- and P-,T-noninvariant corrections to S-matrix
\beq{6.9}
S^{P(PT)}_J(0\oh \to 1\oh)=
\delta S^{(0\,0)}_{P(PT)}+\delta S^{(0\,1p)}_{P(PT)}+
\delta S^{(1s\,0)}_{P(PT)}+\delta S^{(1s\,1p)}_{P(PT)}.
\eeq
The corrections of the first type $\delta S^{(0\,0)}_{P(PT)}$ are
related with a mixing of radial functions of the elastic channel,
$F^{(0)}_s$ and $F^{(0)}_p$, and correspond to direct transition
of a neutron from s- to p-wave due to weak interaction
$\delta\hat{U}_{P(PT)}$ with target nucleons. A target nucleus
remains in a ground state. The corrections of the second type
$\delta S^{(0\,1p)}_{P(PT)}$ are caused by transitions of a
neutron
from the s-wave elastic channel to the p-wave inelastic one with
an excitation of a target to the state with an energy $\ve_{1p}$.
Similarly the corrections of the third type
$\delta S^{(1s\,0)}_{P(PT)}$ correspond to neutron transitions
from the inelastic s-wave channel to the elastic p-wave one, when
a target passes from an excited level with the energy $\ve_{1s}$
to a ground state. Finally, the corrections of fourth type
$\delta S^{(1s\,1p)}_{P(PT)}$ correspond to a mixing of
s- and p-wave functions of the inelastic channels,
$F^{(1)}_s$ and $F^{(1)}_p$,
with a transition of a target from an excited level
$\ve_{1s}$ to another $\ve_{1p}$ one.

In the framework of the model considered we have for these P- and
P-,T-noninvariant corrections
\begin{eqnarray}
&&\delta S^{(0\,0)}_{P(PT)}=\xi_{P(PT)}
\frac{U^{(0\,0)}_{P(PT)}}{\om}
\Bigl( A_sA_pf_{P(PT)}(\kp_s,\kp_p)+
A_sA'_pf_{P(PT)}(\kp_s,\kp'_p)+{}
\nonumber\\[\medskipamount]
&&\phantom{\delta S^{(0\,0)}_{P(PT)}}+
A'_sA_pf_{P(PT)}(\kp'_s,\kp_p)+
A'_sA'_pf_{P(PT)}(\kp'_s,\kp'_p)\Bigr),
\label{6.10}
\end{eqnarray}
\beq{6.11}
\delta S^{(0\,1p)}_{P(PT)}=\xi_{P(PT)}
\frac{U^{(0\,1p)}_{P(PT)}}{\om}
\Bigl( A_sB'_pf_{P(PT)}(\kp_s,\kp'_p)+
A'_sB'_pf_{P(PT)}(\kp'_s,\kp'_p)\Bigr),
\eeq
\beq{6.12}
\delta S^{(1s\,0)}_{P(PT)}=\xi_{P(PT)}
\frac{U^{(1s\,0)}_{P(PT)}}{\om}
\Bigl( B'_sA_pf_{P(PT)}(\kp'_s,\kp_p)+
B'_sA'_pf_{P(PT)}(\kp'_s,\kp'_p)\Bigr),
\eeq
\beq{6.13}
\delta S^{(1s\,1p)}_{P(PT)}=\xi_{P(PT)}
\frac{U^{(1s\,1p)}_{P(PT)}}{\om}
B'_sB'_pf_{P(PT)}(\kp'_s,\kp'_p),
\eeq
where $\xi_P=i$, $\xi_{PT} =1$ and the following designations for
integrals are used
\beq{6.14}
f_P(\kp_s,\kp_p)=
2 \int\limits_0^R dr \sin \kp_sr \; \kp_p \sin \kp_pr-
\sin \kp_sR \left(\frac{\sin \kp_pR}{\kp_pR}-\cos \kp_pR \right),
\eeq
\beq{6.15}
f_{PT}(\kp_s,\kp_p)=\frac{2}{R^2} \int\limits_0^R r dr
\sin \kp_sr \left( \frac{\sin \kp_pr}{\kp_pr}-\cos \kp_pr\right).
\eeq
These integrals are easily calculated
\begin{eqnarray}
&&{\displaystyle f_P(\kp_s,\kp_p)=
\kp_pR \left( \frac{\sin (\kp_s-\kp_p)R}{(\kp_s-\kp_p)R}-
\frac{\sin (\kp_s+\kp_p)R}{(\kp_s+\kp_p)R} \right)-{}}
\nonumber\\[\medskipamount]
&&\qquad \qquad {\displaystyle {}-
\sin \kp_sR\left(\frac{\sin \kp_pR}{\kp_pR}-\cos \kp_pR\right)},
\label{6.16}
\end{eqnarray}
\begin{eqnarray}
&&{\displaystyle f_{PT}(\kp_s,\kp_p)=\frac{1}{\kp_pR}
\left( \frac{\sin (\kp_s-\kp_p)R}{(\kp_s-\kp_p)R}-
\frac{\sin (\kp_s+\kp_p)R}{(\kp_s+\kp_p)R} \right)-{}}
\nonumber\\[\medskipamount]
&&\qquad \qquad {\displaystyle {}-\frac{1}{(\kp_s-\kp_p)R}
\left( \frac{\sin (\kp_s-\kp_p)R}{(\kp_s-\kp_p)R}-
\cos (\kp_s-\kp_p)R \right)-{}}
\nonumber\\[\medskipamount]
&&\qquad \qquad \qquad {\displaystyle
{}-\frac{1}{(\kp_s+\kp_p)R}
\left( \frac{\sin (\kp_s+\kp_p)R}{(\kp_s+\kp_p)R}-
\cos (\kp_s+\kp_p)R \right)}. \label{6.17}
\end{eqnarray}

The depths of P- and T-noninvariant potentials $U^{(ii')}_{P(PT)}$
are external parameters of the model. However to compare the
contributions (\ref{6.10})-(\ref{6.13}) to the observables we need
some estimates for these depths. We assume, that P- or
P-,T-noninvariant operator of neutron-nucleons interaction
$\delta\hat{U}_{P(PT)}$ is a sum of pairwise operators. Then the
functions $U^{(ii')}_{P(PT)}(r)$ on the right parts of
Eqs.(\ref{4.28})-(\ref{4.29}) are single-particle matrix elements.
In the case of parity nonconserving interaction a diagonal
single-particle matrix element is usually evaluated by
\beq{6.18}
U^{(0\,0)}_P \sim \mathop{<\hat{u}_P>}\nolimits_{s.p.} \sim
Gm^2_{\pi}\Omega \sim 0.1 \enskip \mbox{eV},
\eeq
where $G=10^{-5}/m^2_p$ is a Fermi constant, $m_{\pi}$ is a pion
mass, and $\Omega \sim 1$~MeV is a characteristic one-particle
energy. In the case of P-,T-noninvariant interaction we assume,
that
\beq{6.19}
U^{(0\,0)}_{PT} \sim \mathop{<\hat{u}_{PT}>}\nolimits_{s.p.} \sim
\phi Gm^2_{\pi}\Omega,
\eeq
where $\phi < 10^{-3}$ according the present data.

To estimate nondiagonal single-particle matrix elements we use a
well known procedure \cite{5a}-\cite{5c}. We present a target wave
function, which describes a highly excited state, as a
superposition of many simple configurations
\beq{6.20}
\psi_{\alpha}=\sum_{k=1}^N c_k \psi_k.
\eeq
Here $|c_k|\sim 1/N^{1/2}$ owing to normalization condition. A
number $N$ is usually evaluated as $N\sim\om /D$, where $D$ is a
mean distance between excited levels. In this sum few
configurations differ from a ground state $\psi_0$ by an
excitation of one particle only. Therefore
\beq{6.21}
U^{(0\,1p)}_{P(PT)} \sim U^{(1s\,0)}_{P(PT)} \sim
<\alpha |\hat{u}_{P(PT)}|0> \sim
\frac{1}{N^{1/2}}\mathop{<\hat{u}_{P(PT)}>}\nolimits_{s.p.} \sim
\frac{U^{(0\,0)}_{P(PT)}}{N^{1/2}}.
\eeq
Similarly we obtain
\beq{6.22}
U^{(1s\,1p)}_{P(PT)} \sim
<\alpha |\hat{u}_{P(PT)}|\alpha'> \sim
\frac{1}{N}\:N^{1/2}
\mathop{<\hat{u}_{P(PT)}>}\nolimits_{s.p.} \sim
\frac{U^{(0\,0)}_{P(PT)}}{N^{1/2}},
\eeq
where, as usual, a noncoherent sum of $\sim N$ single-particle
matrix elements between simple configurations is estimated by
$N^{1/2}\mathop{<\hat{u}> }\nolimits_{s.p.}$.

We notice, that a mean distance between levels $D_l$ in the
partial wave $l$ and a characteristic reduced width $\gamma_l$ of
resonances are related by order of value \cite{24a}
\beq{6.23}
\gamma_l \sim \frac{D_l}{\pi \kp_l R}.
\eeq
Therefore
\beq{6.24}
\frac{1}{N^{1/2}} \sim \left(\frac{D_l}{\om}\right)^{1/2} \sim
(\pi \kp_lR)^{1/2}\left(\frac{\gamma_l}{\om}\right)^{1/2}.
\eeq
We see, that an enhancement $\sim (\om /\gamma_l)^{1/2}$ of the
factors $B'_l$ (\ref{5.46}) in comparison with $A'_l$ is exactly
canceled by a depression $\sim (\gamma_l/\om )^{1/2}$ of
nondiagonal potentials $U^{(0\,1p)}_{P(PT)}$ and
$U^{(1s\,0)}_{P(PT)}$ in comparison with a diagonal one
$U^{(0\,0)}_{P(PT)}$. This means, that the contributions
(\ref{6.11}), (\ref{6.12}) from a mixing of functions of the
elastic and inelastic channels either less, or considerably do not
surpass the contribution (\ref{6.10}) from a mixing in the elastic
channel! Therefore we shall not discuss the components, which are
proportional to $\delta S^{(0\,1p)}_{P(PT)}$ and
$\delta S^{(1s\,0)}_{P(PT)}$.

We turn now to an analysis and comparison of contributions
$\delta S^{(0\,0)}_{P(PT)}$ and $\delta S^{(1s\,1p)}_{P(PT)}$ to
the observables. In the region far away from both mixing
s- and p-wave resonances ($|\ve-E_l|>\gamma^{1/2}_l\om^{1/2}$), as
we have just established (see Eqs.(\ref{5.52}), (\ref{5.53})), the
components with amplitude $A_l$ dominate in the wave function. In
this region of purely potential scattering only the quantity
\beq{6.25}
\mathop{\left(\delta S^{(0\,0)}_{P(PT)}\right)}\nolimits_{pot}
\sim \xi_{P(PT)} \exp(i(\delta_s+\delta_p))
\frac{U^{(0\,0)}_{P(PT)}}{\om}(kR)^2
\eeq
is significant. We have taken into account, that
$\mathop{(A_l)}\nolimits_{pot}\simeq \exp (i\delta_l)s^{1/2}_l$.
We note, that according to Eqs.(\ref{5.50}) and (\ref{5.60}) we
are dealing here with the total phase shifts $\delta_l$ for
elastic scattering. We get for P- and T-noninvariant observables
\beq{6.26}
\frac{\ep^{(0\,0)}_{P(PT)}}{n}
\sim
\frac{\chi^{(0\,0)}_{P(PT)}}{n}
\sim 4\pi R^2\frac{U^{(0\,0)}_{P(PT)}}{\om}.
\eeq

In the region $|\ve-E_l|<\gamma_l^{1/2}\om^{1/2}$ the factors
$B'_l$ (\ref{5.46}) become considerable. We obtain the following
expression for mixing amplitude in the inelastic channel
(\ref{6.13})
\begin{eqnarray}
&&\delta S^{(1s\,1p)}_{P(PT)}=
\xi_{P(PT)}\exp(i(\delta^c_s+\delta^c_p))
\frac{U^{sp}_{P(PT)}}{\om}
\frac{\om}{(\gamma_s\gamma_p)^{1/2}}(kR)^2\times{}\nonumber
\\[\medskipamount]
&&{}\times\frac{\gamma_s\gamma_p}
{\left(\ve-E_s+is_0
\left(\gamma_s-\frac{\tst \ve-E_s}{\tst \Phi^0_s}\right)\right)
\left(\ve-E_p+is_1
\left(\gamma_p-\frac{\tst \ve-E_p}{\tst \Phi^0_p}\right)\right)},
\label{6.27}
\end{eqnarray}
where
\beq{6.28}
U^{sp}_{P(PT)}=U^{(1s\,1p)}_{P(PT)}
\frac{x_sx_p}{y_sy_p}
\mathop{\left(\frac{f_{P(PT)}(\kp'_s,\kp'_p)}
{\kp'_sR \cos\kp'_sR\:\kp'_pR\sin\kp'_pR}\right)}\nolimits_{E_l}.
\eeq
As far as $|E_s-E_p|\ll\om$, it is unimportant, at what
energy, $E_s$ or $E_p$, the right part is taken here. A general
expression for an amplitude of mixing in the elastic channel
(\ref{6.10}) is rather cumbersome. However, in some special
interesting cases this general expression may be simplified.

We consider at first the region, lying near the resonances, but
beyond their reduced widths
($\gamma_l < |\ve-E_l| < \gamma^{1/2}_l\om^{1/2}$). In particular,
an interval between s- and p-wave resonances can belong to this
region. The factors $A_l$ still dominate over $A'_l$ (see
Eq.(\ref{5.52})). The estimates for them do not change, so the
contributions from mixing in the elastic channel remain on the
level of (\ref{6.25}), (\ref{6.26}). At the same time a mixing in
the inelastic channel gives
\beq{6.29}
\frac{\ep^{(1s\,1p)}_{P(PT)}}{n} \sim
\frac{\chi^{(1s\,1p)}_{P(PT)}}{n} \sim
4\pi R^2\frac{U^{sp}_{P(PT)}}{\om}
\frac{\om \gamma^{1/2}_s \gamma^{1/2}_p}
{(\ve-E_s)(\ve-E_p)}.
\eeq
Though $\om^{1/2}\gamma^{1/2}_l/|\ve-E_l|>1$, the potentials
$U^{sp}_{P(PT)}$ are suppressed in comparison with
$U^{(0\,0)}_{P(PT)}$ by the factor of the scale of (\ref{6.24}).
An approximate evaluation shows, that only in the region
$\gamma_l < |\ve-E_l| < \gamma^{3/4}_l\om^{1/4}$ a mixing in the
inelastic channel provides an increase of P- and T-noninvariant
effects over the background values (\ref{6.26}).

We turn now to an analysis of nearest vicinities of resonances. We
assume, that a distance between s- and p-wave resonances surpasses
their reduced widths. To be more specific,
\beq{6.30}
\gamma_s|\Phi^0_s| \sim
\gamma_p|\Phi^0_p| \ll |E_s-E_p|,
\eeq
so the condition (\ref{5.52}) may be used. Consequently near
p-wave resonance \mbox{($|\ve-E_p|\ll\gamma_p|\Phi^0_p|$)} we have
$|A'_s|\ll |A_s|$. We get for P- (or P-,T-) noninvariant amplitude
of mixing in the elastic channel
\beq{6.31}
\delta S^{(0\,0)}_{P(PT)} \simeq
\xi_{P(PT)}\exp(i(\delta_s+\delta^c_p))
\frac{U^p_{P(PT)}}{\om}(kR)^2
\frac{\gamma_p}{\ve-E_p+i\frac{\tst \Gamma_p}{\tst 2}}.
\eeq
Similarly, near s-wave resonance ($|\ve-E_s|\ll
\gamma_s|\Phi^0_s|$) $|A'_p|\ll |A_p|$, so an amplitude of
interest is of the form
\beq{6.32}
\delta S^{(0\,0)}_{P(PT)} \simeq
\xi_{P(PT)}\exp(i(\delta^c_s+\delta_p))
\frac{U^s_{P(PT)}}{\om}(kR)^2
\frac{\gamma_s}{\ve-E_s+i\frac{\tst \Gamma_s}{\tst 2}}.
\eeq
For simplification we use designations for the depths of
P- and T-noninvariant potentials
\beq{6.33}
U^p_{P(PT)}=U^{(0\,0)}_{P(PT)}x_p
\mathop{\left(\frac{1}{\kp_sR\cos \kp_sR}\right)}\nolimits_{E_s}
\mathop{\left(\frac{f_{P(PT)}(\kp_s,\kp'_p)}
{\kp'_pR\sin \kp'_pR}-
\frac{f_{P(PT)}(\kp_s,\kp_p)}
{\kp_pR\sin \kp_pR}\right)}\nolimits_{E_p},
\eeq
\beq{6.34}
U^s_{P(PT)}=U^{(0\,0)}_{P(PT)}x_s
\mathop{\left(\frac{1}{\kp_pR\cos \kp_pR}\right)}\nolimits_{E_p}
\mathop{\left(\frac{f_{P(PT)}(\kp'_s,\kp_p)}
{\kp'_sR\cos \kp'_sR}-
\frac{f_{P(PT)}(\kp_s,\kp_p)}
{\kp_sR\cos \kp_sR}\right)}\nolimits_{E_s},
\eeq
just as in Eqs.(\ref{6.27}), (\ref{6.28}).

Let us compare the amplitudes in resonances with the background
value of (\ref{6.25}). If $\ve=E_l$, the amplitudes (\ref{6.31})
and (\ref{6.32}) of s- and p-wave mixing in the elastic channel
surpass the potential value (\ref{6.25}) by the factor of
$\sim 1/s_l$. The reason is that in the potential scattering a
jump of potential and centrifugal barrier hinder a penetration of
a neutron into a nucleus, while in the resonance these factors
hinder an exit of a neutron from a nucleus, increasing an
interaction time and, therefore, a mixing amplitude (see
Eqs.(\ref{5.55}), (\ref{5.56}) and (\ref{5.58})). We consider now
the amplitude of s- and p-wave mixing in the inelastic channel
(\ref{6.27}). In the resonance ($\ve=E_l$) it is enhanced by the
factor $\om /\Gamma_l$. According to Eq.(\ref{5.59}) this quantity
corresponds to the delay of a neutron inside a nucleus due to a
weak coupling between the channels. However, account must be taken
of the suppression of the depth of P- (P-,T-) noninvariant
potential $U^{sp}_{P(PT)}$ in comparison with $U^{(0\,0)}_{P(PT)}$
by the factor of $1/N^{1/2}\sim (\gamma_l/\om)^{1/2}$.
Nevertheless, multiplying these factors, we obtain in the
inelastic channel an enhancement
$(1/s^{1/2}_l)(\om/\Gamma_l)^{1/2}\sim
(1/s_l)(\om/\gamma_l)^{1/2}$. This quantity surpasses an
enhancement in the elastic channel by the factor of
$\sim (\om/\gamma_l)^{1/2}$!

Thus, far away from s- and p-wave resonances
($|\ve-E_l|>\gamma^{3/4}_l\om^{1/4}$) P- and T-noninvariant
effects are determined by a mixing in the elastic channel (by the
quantity $\delta S^{(0\,0)}_{P(PT)}$). In a resonance the
quantities $\delta S^{(0\,0)}_{P(PT)}$ and
$\delta S^{(1s\,1p)}_{P(PT)}$ peak. However,
$\delta S^{(1s\,1p)}_{P(PT)}$ surpasses
$\delta S^{(0\,0)}_{P(PT)}$ by the factor of
$\sim (\om/\gamma_l)^{1/2}$. Therefore, P- and T-noninvariant
effects in resonances result mainly from a mixing in the
inelastic channel.

For the first time an existence of an enhancement factor
$(\om/\Gamma)^{1/2}$ in the region of compound resonances was
mentioned in Refs.\cite{11a,11b} in connection with an analysis of
sensitivity of detailed balance tests to T-invariance violation.
In resonances the time of neutron-nucleus interaction increases;
so the P- and T-noninvariant mixing amplitudes are enhanced. In
such
context the form $(\om/\Gamma)^{1/2}$ of an enhancement factor
seems natural.

However, the other form of an enhancement factor is usually used.
One separates the ratio $U^{sp}_{P(PT)}/(E_p-E_s)$ in the
amplitude of mixing in the inelastic channel (\ref{6.27}). This
quantity is interpreted as the amplitude of P- or
P-,T-noninvariant mixing of s-wave resonance to p-wave one or vice
versa. As far as $|E_p-E_s|\sim D\sim \om/N$, and
$U^{sp}_{P(PT)}\sim U^{(0\,0)}_{P(PT)}/N^{1/2}$, we have
\beq{6.35}
\frac{U^{sp}_{P(PT)}}{|E_p-E_s|}\sim
N^{1/2}\frac{U^{(0\,0)}_{P(PT)}}{\om}.
\eeq
Such enhancement is said to be dynamic. It is related with a
proximity of mixing resonances. For the first time it was analyzed
in Refs.\cite{5a}-\cite{5c}.

The amplitude of s- and p-wave mixing in compound resonance is
usually taken in the form
\beq{6.36}
\delta S^{sp}_{P(PT)}=\frac{\xi_{P(PT)}}{2}
\exp(i(\delta^c_s+\delta^c_p))
U^{sp}_{P(PT)}
\frac{(\Gamma^n_s)^{1/2}(\Gamma^n_p)^{1/2}}
{\left(\ve-E_s+i\frac{\tst \Gamma_s}{\tst 2}\right)
\left(\ve-E_p+i\frac{\tst \Gamma_p}{\tst 2}\right)}.
\eeq
With regard to coincidence in the model of neutron width,
$\Gamma^n_l$,
and total width, $\Gamma_l$, this expression is very
similar to Eq.(\ref{6.27}), but is not identical to it. The reason
is that the denominators $[l]$ (\ref{5.49}), entering in the
formula
(\ref{6.27}), have usual Breit-Wigner form only in the vicinities
of resonance energies, which are small in comparison with reduced
widths. As far as the distances between s- and p-wave resonances
are usually comparable or surpass the reduced widths of
resonances, the denominator $[s]$ has not the Breit-Wigner form
near p-wave resonance, while a similar deviation takes place for
the quantity $[p]$ near s-wave resonance. These deviations are,
however, rather simple. If $|\ve-E_l|\gg\gamma_l|\Phi^0_l|$, then
we get according to Eq.(\ref{5.60})
\beq{6.37}
\exp(i\delta^c_l)\frac{1}{[l]}\simeq
\exp(i\delta_l)\frac{1}{\ve-E_l}.
\eeq
Thus, the amplitude of mixing in the inelastic channel
(\ref{6.27}) near p-wave resonance is of the form
\beq{6.38}
\delta S^{(1s\,1p)}_{P(PT)}\simeq
\frac{\xi_{P(PT)}}{2}\exp(i(\delta_s+\delta^c_p))
U^{sp}_{P(PT)}
\frac{(\Gamma^n_s)^{1/2}(\Gamma^n_p)^{1/2}}
{\tst \left(\ve-E_s\right)
\left(\ve-E_p+is_1
\left(\gamma_p-\frac{\tst \ve-E_p}{\tst \Phi^0_p}\right)\right)},
\eeq
while near s-wave resonance we obtain
\beq{6.39}
\delta S^{(1s\,1p)}_{P(PT)}\simeq
\frac{\xi_{P(PT)}}{2}\exp(i(\delta^c_s+\delta_p))
U^{sp}_{P(PT)}
\frac{(\Gamma^n_s)^{1/2}(\Gamma^n_p)^{1/2}}
{\tst \left(\ve-E_s+is_0
\left(\gamma_s-\frac{\tst \ve-E_s}{\tst \Phi^0_s}\right)\right)
\left(\ve-E_p\right)}.
\eeq

General expressions for the energy dependent P- and T-noninvariant
observables (\ref{6.2}), (\ref{6.4}), caused by a mixing in the
inelastic channel, are rather cumbersome. However, the situation
is simplified close to resonance energies due to approximate
relationships (\ref{6.38}), (\ref{6.39}). Near p-wave resonance we
have with an accuracy of $(kR)^2$
\beq{6.40}
\frac{\ep^p_{P(PT)}}{n}=
-\nu_{P(PT)}\frac{2\pi}{k^2}g_JU^{sp}_{P(PT)}
\frac{kR\left(1-\frac{\tst 1}{\tst \Phi^0_s}\right)
(\Gamma^n_s)^{1/2}(\Gamma^n_p)^{1/2}
\left(\ve-E_p+
\frac{\tst (kR)^2\gamma_p}
{\tst 1-\frac{\tst 1}{\tst \Phi^0_s}}\right)}
{\left(\ve-E_s\right)
\left((\ve-E_p)^2+(s_1)^2
\left(\gamma_p-
\frac{\tst \ve-E_p}{\tst \Phi^0_p}\right)^2\right)},
\eeq
\beq{6.41}
\frac{\chi^p_{P(PT)}}{n}=
\nu_{P(PT)}\frac{2\pi}{k^2}g_JU^{sp}_{P(PT)}
\frac{(\Gamma^n_s)^{1/2}(\Gamma^n_p)^{1/2}(\ve-E_p)}
{\left(\ve-E_s\right)
\left((\ve-E_p)^2+(s_1)^2
\left(\gamma_p-
\frac{\tst \ve-E_p}{\tst \Phi^0_p}\right)^2\right)},
\eeq
where $\nu_P=1$, $\nu_{PT}=-C_{J\oh}/g_J$. Similarly near s-wave
resonance P- and T-noninvariant observables take the form
\beq{6.42}
\frac{\ep^s_{P(PT)}}{n}=
-\nu_{P(PT)}\frac{2\pi}{k^2}g_JU^{sp}_{P(PT)}
\frac{kR\left(1-\frac{\tst 1}{\tst \Phi^0_s}\right)
(\Gamma^n_s)^{1/2}(\Gamma^n_p)^{1/2}
\left(\ve-E_s+
\frac{\tst \gamma_s}
{\tst 1-\frac{\tst 1}{\tst \Phi^0_s}}\right)}
{\left(\ve-E_p\right)
\left((\ve-E_s)^2+(s_0)^2
\left(\gamma_s-
\frac{\tst \ve-E_s}{\tst \Phi^0_s}\right)^2\right)},
\eeq
\beq{6.43}
\frac{\chi^s_{P(PT)}}{n}=
\nu_{P(PT)}\frac{2\pi}{k^2}g_JU^{sp}_{P(PT)}
\frac{(\Gamma^n_s)^{1/2}(\Gamma^n_p)^{1/2}
(\ve-E_s-(kR)^2\gamma_s)}
{\left(\ve-E_p\right)
\left((\ve-E_s)^2+(s_0)^2
\left(\gamma_s-
\frac{\tst \ve-E_s}{\tst \Phi^0_s}\right)^2\right)}.
\eeq
We see, that in the vicinities of resonances the transmission
asymmetries $\ep^l_{P(PT)}$ and angles of spin rotation
$\chi^l_{P(PT)}$ cross zero. However, it is easy to establish,
that in the intervals of energy
\beq{6.44}
|\ve-E_l|\le\frac{\Gamma_l}{2}=(kR)^{2l+1}\gamma_l
\eeq
the quantities $\ep^l_{P(PT)}$ peak and may be presented in the
form
\beq{6.45}
\frac{\ep^l_{P(PT)}}{n}=
-{\cal P}^l_{P(PT)}(\ve)\si_l(\ve).
\eeq
Here $\si_l(\ve)$ is a partial cross section of neutron-nucleus
interaction
\beq{6.46}
\si_l(\ve)=\frac{\pi}{k^2}g_J
\frac{\Gamma^n_l \Gamma_l}
{(\ve-E_l)^2+\left(\frac{\tst \Gamma_l}{\tst 2}\right)^2},
\eeq
while factors ${\cal P}^l_{P(PT)}(\ve)$ are slow functions of
energy
\beq{6.47}
{\cal P}^p_{P(PT)}(\ve)=
\nu_{P(PT)}\frac{U^{sp}_{P(PT)}}{\ve-E_s}
\left(\frac{\Gamma^n_s}{\Gamma^n_p}\right)^{1/2},
\eeq
\beq{6.48}
{\cal P}^s_{P(PT)}(\ve)=
\nu_{P(PT)}\frac{U^{sp}_{P(PT)}}{\ve-E_p}
\left(\frac{\Gamma^n_p}{\Gamma^n_s}\right)^{1/2}.
\eeq
Transmission asymmetry $\ep^l_{P(PT)}$ crosses zero outside an
interval (\ref{6.44}). While an angle $\chi^l_{P(PT)}$
crosses zero within the width of resonance $\Gamma_l$. Neglecting
a shift $(kR)^2\gamma_s$ in comparison with a width
$\Gamma_s/2=(kR)\gamma_s$, we obtain for the quantities
$\chi^l_{P(PT)}$ in the intervals (\ref{6.44})
\beq{6.49}
\chi^l_{P(PT)}=
-\frac{2(\ve-E_l)}{\Gamma_l}\ep^l_{P(PT)}.
\eeq
The resonance formulas (\ref{6.45})-(\ref{6.49}) coincide with
those found by using a mixing amplitude in the standard form
(\ref{6.36}).

In experiments one chooses a thickness of a target to provide an
optimum rate of statistic gathering. In a resonance $\ve\sim E_l$
this condition gives (see, for example, Ref.\cite{4a})
\beq{6.50}
n\sim \frac{1}{\si_l(E_l)}.
\eeq
It follows that the quantities ${\cal P}^l_{P(PT)}(E_l)$
approximately equal to measured asymmetries $\ep^l_{P(PT)}$ at the
energies $\ve=E_l$ and angles $\chi^l_{P(PT)}$ at the energies
$\ve=E_l\pm\Gamma_l/2$.

We see now, that in a p-wave resonance a measured quantity
(\ref{6.47}) is enhanced by a factor of
$(\Gamma^n_s/\Gamma^n_p)^{1/2})\sim 1/kR$ besides the factor
(\ref{6.35}). This enhancement is said to be structural or
kinematic. While in a s-wave resonance (see
Eq.(\ref{6.48})) a factor of suppression
$(\Gamma^n_p/\Gamma^n_s)^{1/2})\sim kR$ arises. We remind, that in
the model considered a total width of resonance $\Gamma_l$
coincides with its neutron width $\Gamma^n_l$. This situation
approximately corresponds to compound resonances in light nuclei.
Usually the Eqs.(\ref{6.45})-(\ref{6.49}) are analyzed as applied
to heavy nuclei, where $\Gamma_l\gg\Gamma^n_l$, and
$\Gamma_s\sim\Gamma_p$. Let us show, that the additional factors
of enhancement or suppression, arising in Eqs.(\ref{6.47}) and
(\ref{6.48}), have different meaning for light and heavy nuclei.

Let $\Gamma_s=\Gamma^n_s \gg \Gamma_p=\Gamma^n_p$. Thus it is easy
to see, that s- and p-wave resonance cross sections (\ref{6.46})
reach approximately equal maximal values. Therefore, according to
condition (\ref{6.50}) measurements in s- and p-wave resonances
should be carried on targets with the same thickness. But as far
as the quantity ${\cal P}^p_{P(PT)}(E_p)$ is enhanced in
comparison with ${\cal P}^s_{P(PT)}(E_s)$ by a factor of
$\sim 1/(kR)^2$, the P- and T-noninvariant observables
(\ref{6.45}), (\ref{6.49}) will be a $\sim 1/(kR)^2$ times more in
the p-wave resonance than in the s-wave one. It has been just this
result that was obtained earlier in the model. Indeed, an
enhancement factor of the amplitude $\delta S^{(1 s\,1p)}_{P(PT)}$
in a resonance in the partial wave $l$ was found equal
$(1/s_l)(\om /\gamma_l)^{1/2}$. Clearly, this factor is a
$s_0/s_1\sim 1/(kR)^2$ times more in the p-wave than in the
s-wave.

We consider now a case
$\Gamma_s \sim \Gamma_p \gg \Gamma^n_s \gg \Gamma^n_p$. A
resonance cross section $\si_s(E_s)$ surpasses now $\si_p(E_p)$ by
a factor of $\sim 1/(kR)^2$. But the quantity
$\mbox{\cal P}^s_{P(PT)}(E_s)$ is a $\sim 1/(kR)^2$ times less
than $\mbox{\cal P}^p_{P(PT)}( E_p)$. Therefore, in heavy nuclei
one should expect the same P- and T-noninvariant effects
(\ref{6.45}), (\ref{6.49}) in the s- and p-wave resonances,
if the targets with equal thickness are used. However,
in view of condition (\ref{6.50}), in the p-wave resonance
measurement a target should be used a $\sim 1/(kR)^2$ times
thicker, than in the s-wave resonance measurement. Clearly, in
this situation the observables in the p-wave resonance appear
a $\sim 1/(kR)^2$ times more, than in the s-wave resonance.

Thus, in light nuclei an additional enhancement of
P- and T-noninvariant effects in p-wave resonance by a factor of
$(\Gamma^n_s/\Gamma^n_p)^{1/2}\sim 1/kR$ results from an increase
of neutron delay inside a nucleus. In heavy nuclei the total
widths of s- and p-wave resonances are determined by radiative
transitions, therefore the life times of these resonances are
approximately equal. So an additional enhancement in the p-wave
resonance, having the same value
$(\Gamma^n_s/\Gamma^n_p)^{1/2}\sim 1/kR$ that in light nuclei,
arises rather from the advantages of use of thick target.

\section{Model calculation of P- and T-noninvariant effects}
\label{s7}

In the previous section we have shown that the inelastic channel
contributes decisively to the mixing of s- and p-waves. Within the
widths of resonances we have got the usual expressions
(\ref{6.45})--(\ref{6.49}) for P- and T-noninvariant observables.
As an illustration let us present the results of exact calculation
of P- and T-noninvariant effects (\ref{6.2}), (\ref{6.4}) for a
close-lying pair of s- and p-wave resonances.

As far as the model does not account radiative channels, it seems
reasonable  to use it for description of resonances in
light nuclei. We take as an example a pair of s- and p-wave
resonances of the $^{35}$Cl nucleus with energies $E_s=26.60$ keV
and $E_p=22.41$ keV and neutron widths $2g\Gamma^n_s=130$ eV and
$2g\Gamma^n_p=4$ eV \cite{25a}. Radiative contributions to these
widths are $\sim 0.5$ eV only; we neglect them. The nucleus
$^{35}$Cl has spin and parity $I^{\pi}=3/2^+$. We assume that
spins of the resonances chosen equal $J=1$ (there is no
information about these spins in Ref.\cite{25a}).

We remind, that the partial wave is specified by three quantum
numbers $l$, $j$ and $J$. We restrict our attention to two partial
waves $l=0$, $j=1/2$, $J=1$ and $l=1$, $j=1/2$, $J=1$. For brevity
we shall label the quantities, associated with these waves, only
by indexes s and p.

Neutron-nucleus potential in the elastic and inelastic channels is
taken to be spherical potential well of the radius $R=6.5$ fm and
depth
\beq{7.1}
U=U_0+U_{ls}\,\hat{\vec{l}}\,\hat{\vec{s}},
\eeq
where $U_0=23$ MeV and $U_{ls}=1$ MeV. So the parameters of
potential (\ref{5.6}) for s1/2- and p1/2-waves equal $U_s=23$ MeV
and $U_p=22$ MeV. Characteristic one-particle energy (\ref{5.4})
is $\om=490.444$ keV. The sequence of bound states in the
potential chosen is presented in Fig.1. We assume that 18 neutrons
of $^{35}$Cl nucleus fill the levels 1s1/2, 1p3/2, 1p1/2, 1d5/2
and 1d3/2 so the states 2s1/2 and 2p1/2 are free. In the
framework of the model, described in the section \ref{s5},
s- and p-wave resonances correspond to transitions of the incident
neutron to bound states 2s1/2 and 2p1/2, respectively, accompanied
by excitation of the target to the levels with energies $\ve_s$
and $\ve_p$.

Using Eq.(\ref{5.34}) we have found the reduced widths of
resonances
$\gamma_s=372$ eV and $\gamma_p=285$ eV. We have in each partial
wave the set of two equations
\beq{7.2}
\left\{
\begin{array}{l}
\Phi_l(E_l)=0,\\[\bigskipamount]
\mathop{\frac{\tst d}{\tst d \ve}
\Phi_l(\ve)}\nolimits_{\scriptstyle |\ve=E_l}=
-\frac{\tst 1}{\tst \gamma_l},
\end{array}
\right.
\eeq
that gives us two parameters of the model: $\ve_l$ and $W_l$.
Solving these sets we obtain $\ve_s=8831.277$~keV,
$W_s=174.509$~keV and $\ve_p=1589.982$~keV, $W_p=54.422$~keV. We
notice, that dimensionless parameter $\lal$ takes here the values
$\lambda_s=0.020$ and $\lambda_p=0.034$, so we are dealing with
the case of weak coupling between channels ($\lal \ll 1$).

A weak coupling approximation was studied in the section
\ref{s52}. Its quality may be checked by calculation of reduced
widths using Eq.(\ref{5.41}). We get then for the s-wave width
372~eV,
differing from the exact value only in the first decimal, while
for the p-wave width --- 294~eV, differing from the exact value
by~3~\%.

In Fig.2 the energy dependencies of logarithmic derivatives
$\Phi_l(\ve)$ (\ref{5.35}) are shown. Their forms agree well with
Eq.(\ref{5.40}). It is seen that the Breit-Wigner approximation
(\ref{5.32}) holds only in a small vicinity of each resonance. Far
from the resonances the functions $\Phi_l(\ve)$ reach the
potential values (\ref{5.36}), which equal $\Phi^0_s=10.706$ and
$\Phi^0_p=-3.188$.

Using Eq.(\ref{5.27}) it is easy to calculate the phase shifts for
each partial wave
\beq{7.3}
\delta_l(\ve)=\delta^c_l(\ve)+
\arctg \frac{s_l(\ve)}{\Phi_l(\ve)},
\eeq
and cross sections
\beq{7.4}
\si_l(\ve)=\frac{4\pi}{k^2}g_J \sin^2\delta_l(\ve).
\eeq
They are presented in Fig.3. We have for these cross sections at
the resonance energies $\si_s(E_s)=36.681$~b and
$\si_p(E_p)=43.572$~b.

The amplitudes $A_l$ and $A'_l$ (see Eqs.(\ref{5.25})) are
complex. However, as noted above (see text before
Eq.(\ref{5.60})), at $\ve<\ve_l$ the phases of these factors
coincide with the phase shift $\delta_l$. Let
$A_l=\alpha_l\exp(i\delta_l)$ and $A'_l=\alpha'_l\exp(i\delta_l)$.
In Fig.4 the real quantities $\alpha_l$ and $\alpha'_l$ are shown
versus the energy. According to the Eqs.(\ref{5.44}), (\ref{5.45})
and (\ref{5.56}) an enhancement of the factors $A_l$ and $A'_l$ is
$\sim 1/kR$ times as large for the p-wave resonance as for the
s-wave one.
We have here $k(E_s)R=0.233$ and $k(E_p)R=0.214$, so the
enhancement is on the scale of $\sim$ 4 -- 5. An increase of
the amplitudes near the resonances is displayed in Figs.4a and
4b. Fig.4c shows that the factors $A_l$ fall more slowly than
$A'_l$ (see Eq.(\ref{5.52})). According to Eqs.(\ref{5.15})
to obtain the factors $B'_s$ and $B'_p$ we multiply $A'_s$ and
$A'_p$ by $-1/\lambda_s=-50.626$ and $-1/\lambda_p=-29.250$,
respectively.

We turn now to estimates for P- and T-noninvariant observables
(\ref{6.2}) and (\ref{6.4}). We restrict our attention to the
dominating contribution $\delta S^{(1s\,1p)}_{P(PT)}$ to the sum
(\ref{6.9}). Let
\beq{7.5}
U^{(0\,0)}_P=10^{-1}\mbox{ eV}, \qquad \qquad
U^{(0\,0)}_{PT}=10^{-4}\mbox{ eV},
\eeq
so an unknown constant $\phi$ is taken to be $10^{-3}$. According
to Eqs.(\ref{6.22}) and (\ref{6.24}) we take as the depth of a
nondiagonal weak potential the quantity
\beq{7.6}
U^{(1s\,1p)}_P=
\mathop{\left(\frac{\pi \kp_sR\gamma_s}
{\om}\right)}^{1/2}\nolimits_{E_s}U^{(0\,0)}_{P}=
1.3\cdot 10^{-2}\mbox{ eV},
\eeq
similarly
\beq{7.7}
U^{(1s\,1p)}_{PT}=1.3\cdot 10^{-5}\mbox{ eV}.
\eeq
It is easy to estimate, that the potential values (\ref{6.26}) of
P-noninvariant observables are thus on the scale of
$\sim 10^{-6}$~b, while for P-,T-noninvariant ones we get
$\sim 10^{-9}$~b.

Taking into account the identity of the phases of the factors
$B'_s$ and $B'_p$ and the phase shifts for elastic scattering and
putting $B'_l=\beta'_l\exp(i\delta_l)$, we obtain from
Eq.(\ref{6.13}) for the observables
\beq{7.8}
\frac{\ep_{P(PT)}}{n}=
\nu_{P(PT)}\frac{4\pi}{k^2}g_J\sin (\delta_s+\delta_p)
\frac{U^{(1s\,1p)}_{P(PT)}}{\om}
\beta'_s \beta'_p f_{P(PT)}(\kp'_s,\kp'_p),
\eeq
\beq{7.9}
\frac{\chi_{P(PT)}}{n}=
\nu_{P(PT)}\frac{4\pi}{k^2}g_J\cos (\delta_s+\delta_p)
\frac{U^{(1s\,1p)}_{P(PT)}}{\om}
\beta'_s \beta'_p f_{P(PT)}(\kp'_s,\kp'_p).
\eeq
We notice, that for the case considered here $J=I-1/2$ we have
$C_{J\oh}=g_J$, so $\nu_P=1$, and $\nu_{PT}=-1$. The integrals
$f_{P(PT)}(\kp'_s,\kp'_p)$ depend only slightly on the energy
(they belong to the functions, that change significantly only on
the scale of $\om$). At $\ve=E_s$ these integrals equal
$f_P(\kp'_s,\kp'_p)=4.892$ and $f_{PT}(\kp'_s,\kp'_p)=0.522$.

In Figs.5 and 6 the asymmetries $\ep_{P(PT)}/n$ and angles
$\chi_{P(PT)}/n$ are presented versus the energy. Figs.5a, 6a
and 5b, 6b show the effects near p- and s-wave resonances,
respectively. Within the widths of the resonances these curves are
described well by Eqs.(\ref{6.45})-(\ref{6.49}). In particular,
the angles of spin rotation cross zero at $\ve\simeq E_l$. In the
same time in accordance with Eqs.(\ref{6.40}), (\ref{6.42}) the
transmission asymmetries cross zero at the energies
$\ve\simeq E_l-\Gamma_l/(2kR(1-1/\Phi^0_s))$, that is before the
resonances beyond their widths. Figs.5c, 6c display it.

We see that the effects are maximal in the p-wave resonance. The
asymmetry $\ep_P/n$ reach the value $\sim 4\cdot 10^{-2}$~b, which
is $\sim 4\cdot 10^4$ times greater than the potential value! An
enhancement factor involves $(\om/\gamma_p)^{1/2} \sim 40$,
$1/(kR)^3 \sim 10^2$ and numerical coefficients like
$f_P(\kp'_s,\kp'_p)$. While we get
$\sim 4\cdot 10^{-6}$~b for the maximal value of $\ep_{PT}/n$,
which surpasses the potential estimate by a factor
$\sim 4\cdot 10^3$ only. The reason is that the integral of the
overlap $f_{PT}(\kp'_s,\kp'_p)$ is found one order lower
than $f_P(\kp'_s,\kp'_p)$. Clearly, the energy dependencies of
P- and P-,T-noninvariant observables are the same. So the curves
in Figs.5 and 6 differ only in sign and factor of $\sim 10^{-4}$.
So in what follows we discuss the P-nonconserving effects only.

In the resonance measurements the target thicknesses should be
taken to be $n\sim 1/\sigma_l(E_l) \sim 0.02$~b$^{-1}$.
In the p-wave resonance we obtain
for P-noninvariant asymmetry $\ep_P$ and angle $\chi_P$ an
estimate $\sim 10^{-3}$ (for P-,T-noninvariant observables
$\ep_{PT}$ and $\chi_{PT}$ --- $\sim 10^{-7}$). In the s-wave
resonance these quantities are $1/(kR)^2 \sim 25$ times
suppressed.

It is interesting, that there exist the regions beyond
the resonance widths, which are favorable for measurements. This
is because the quantities $\ep_{P(PT)}$ and
$\chi_{P(PT)}$ fall more slowly than the cross sections.
Fig.7 shows, that to the right and
to the left of the p-wave resonance there are the energy
intervals, where the cross section equals $\sim 1$~b, while
$\ep_P/n\sim 0.5\cdot 10^{-3}$~b and $\chi_P/n \sim 10^{-3}$~b. On
these intervals to provide the optimum rate of statistic gathering
the target thickness should be equal to $n\sim 1$~b$^{-1}$, so the
observables $\ep_P$ and $\chi_P$ are of the same scale
$\sim 10^{-3}$, than in the p-wave resonance. We notice, that a
total s-wave cross section near p-wave resonance is a sum of
contributions of partial waves $l=0$, $j=1/2$, $J=1$ and $l=0$,
$j=1/2$, $J=2$. The latter was omitted by us. However, these
contributions are, obviously, comparable. So our conclusions
remain valid.

A completely different type of situation occurs near
an interference minimum to the left of s-wave resonance. In the
case considered, when a target spin $I$ differs from zero, there
are no deep minimums in a total cross section owing to
superposition of the contributions from waves with $J=I\pm 1/2$.
We remind, that the P-,T-noninvariant correlation
$({\bf \sig}[{\bf n}_k{\bf n}_I])$ could be studied only on the
targets with nonzero spin $I$. However, one may observe
P-nonconserving effects in the interaction of neutrons with
spinless target nuclei. In this case only one partial wave $l=0$,
$j=J=1/2$ contributes to s-wave cross section. Therefore, let us
consider a vicinity of an interference minimum, keeping in mind
the possible applications to the case of neutron interaction with
spinless nuclei.

In Fig.8 the calculated quantities $\ep_P/n$, $\chi_P/n$ and
$\sigma(\ve)=\sigma_s(\ve)+\sigma_p(\ve)$ are presented. In
accordance with Fig.3 a minimum of cross section is determined by
p-wave scattering. In fact a radiative capture is of first
importance in the minimum. It is easy to estimate a contribution
from the nearest s-wave resonance, taking its radiative width to
be $\Gamma^{\gamma}_s=0.5$~eV. A capture cross section
\beq{7.10}
\sigma^{\gamma}_s(\ve)=
\frac{\pi}{k^2}g_J
\frac{\Gamma^n_s\Gamma^{\gamma}_s}
{(\ve-E_s)^2+\left(\frac{\tst \Gamma_s}{\tst 2}\right)^2}
\eeq
is found to be 0.013~b in the minimum of the scattering cross
section at $\ve\simeq 26.2$~keV. We notice now, that for the value
$\sim 0.1$~b of the total cross section a target with thickness
$n\sim 10$~b$^{-1}$ gives the optimum rate of
statistic gathering. So the measuring angle of spin rotation in
the region of an interference minimum may have the magnitude
$\chi_P\sim 10^{-3}$, which is comparable with the effect in
the p-wave resonance! A situation with an asymmetry is worse
as this quantity crosses zero in the interval considered.

\section{Summary}
\label{s8}

The paper deals with a simplified model for resonant
neutron-nucleus interaction based on the scheme of coupled
channels. The equations and boundary conditions for radial
functions are written in section \ref{s2}. The form of diagonal
and nondiagonal potentials, as well as a number of channels are
external parameters of the model.

Perturbations violating P- and T-invariance are weak enough to be
taken into account in the first order. The formulas for
P- and T-noninvariant corrections to S-matrix are presented in
section \ref{s4}. The single-particle operators of weak
interaction of the form $\sig{\bf p}$ and $\sig{\bf r}$ are
considered in details.

The analytical solution for the problem with two coupled channels
and square-well potentials is given in section \ref{s5}. A neutron
orbital momentum $l$ is arbitrary. The case of a weak channel
coupling corresponding to narrow Breit-Wigner resonance is
analyzed.

A mixing of two narrow s- and p-wave resonances by P- and
P-,T-noninvariant potentials is studied in section \ref{s6}. In
the framework of the model the contributions of four types to the
mixing amplitude are separated. Here, we are dealing with
transitions from s- to p-wave (or vice versa) in the elastic
channel, inelastic channel, as well as with the cross terms. It is
shown, that the cross terms do not exceed the contribution from
a mixing in the elastic channel. An enhancement of the mixing
amplitude in the elastic channel in a resonance with an orbital
momentum $l$ is of the scale $\sim 1/(kR)^{2l+1}$, while that in
the inelastic channel reaches
$\sim(\om /\gamma_l)^{1/2}(1/(kR)^{2l+1})$.
Here $\om$ is the characteristic one-particle energy and
$\gamma_l$ is the reduced width of the resonance. Thus, the
contribution from a mixing in the inelastic channel dominates.

The analytical expression for the amplitude of mixing in the
inelastic channel differs slightly from the usually used formula
for compound-compound mixing. The reason is the resonant
amplitudes deviate from the Breit-Wigner energy dependence beyond
the widths of resonances. Nevertheless we obtain the usual
expressions for P- and T-noninvariant observables within the
widths of resonances. As far as the total width of a resonance
coincides with the neutron width, the model is directly
appropriate for light nuclei only. In this situation an expansion
of the factor of resonance enhancement into dynamic and structural
(kinematic) ones is conventional.

A close-lying pair of s- and p-wave resonances of the nucleus
$^{35}$Cl is reproduced in section \ref{s7}. We emphasize, that
our description of neutron-nucleus interaction is very schematic.
A resonance enhancement of P- and T-noninvariant observables is
demonstrated. The favorable possibilities are shown to exist on
thick targets for measurements beyond the resonance widths. In
particular, an interference minimum near s-wave resonance is of
interest for P-odd neutron spin rotation on light spinless
nuclei.

An including of radiative channels will allow to analyze
P- and T-noninvariant effects, first, for heavy nuclei, secondly,
in radiative neutron capture. On the other hand, an increase of a
number of coupled channels will lead to large sets of
s- and p-wave resonances. It may be significant for an analysis of
the sign correlation of P-odd effects \cite{10a}. A replacement of
square-well potentials by potentials of Woods-Saxon type, which
are convenient for numerical solution of coupled equations, will
make the model more realistic.
\vspace{\baselineskip}

\noindent{\Large \bf Acknowledgements}
\vspace{\baselineskip}

I wish to thank Yu.V.Gaponov, B.V.Danilin, and N.B.Shul'gina for
useful discussions. This study had partial financial support from
the International Science Foundation, the grant number is M7C000.
\vspace{\baselineskip}
\vspace{\baselineskip}

\appendix

\noindent{\Large \bf Appendix}
\vspace{\baselineskip}

\noindent The numerical factors A, B, and C enter into the
amplitude of elastic scattering of s- and p-wave neutrons by
polarized nuclei (see Eqs.(\ref{3.1}), (\ref{3.4})-(\ref{3.11})).
They are expressed in terms of the normalized Racah functions and
9j-simbols
$$
A^{(1)}_J=-g_J\left(\frac{3I}{I+1}\right)^{1/2}
U(J\oh I1,I\oh), \eqno (A1)
$$
$$
A^{(2)}_{Jjj'}=-g_J\left(\frac{3I}{I+1}\right)^{1/2}
U(JjI1,Ij')U(1\oh j'1,j\oh), \eqno (A2)
$$
$$
A^{(3)}_{Jjj'}=-3g_J\left(\frac{6I}{I+1}\right)^{1/2}
(2j+1)^{1/2}U(JjI1,Ij')
\left\{\begin{array}{ccc}
j'&1&1/2\\
j&1&1/2\\
1&2&1
\end{array}\right\}, \eqno (A3)
$$
$$
B_{Jj}=(-1)^{3/2-j}g_J\left(\frac{3I}{I+1}\right)^{1/2}
U(J\oh I1,Ij), \eqno (A4)
$$
$$
C_{Jj}=3(-1)^{j-1/2}g_J\left(\frac{I}{2(I+1)}\right)^{1/2}
U(J\oh I1,Ij)U(\oh 1j1, \oh 1). \eqno (A5)
$$

Using the explicit expressions for Racah functions \cite{22a} and
9j-simbols \cite{26a}, we get
$$
A^{(1)}_J=-3A^{(2)}_{J\oh\oh}=\tth A^{(3)}_{J\oh\oh}=
B_{J\oh}=-C_{J\oh}, \eqno (A6)
$$
$$
A^{(2)}_{J\oh\th}=A^{(2)}_{J\th\oh}=
4A^{(3)}_{J\oh\th}=4A^{(3)}_{J\th\oh}=
\frac{2}{3}B_{J\th}=\frac{4}{3}C_{J\th}, \eqno (A7)
$$
$$
A^{(2)}_{J\th\th}=-5A^{(2)}_{J\th\th}, \eqno (A8)
$$
where

\begin{center}
$$
\renewcommand{\arraystretch}{0}
\begin{array}{|c|c|c|c|}
\hline
 & \hphantom{\dst -3(2I+1)(I+1)} &
\hphantom{\dst -3(2I+1)(I+1)} &
\hphantom{\dst -3(2I+1)(I+1)} \\
\vph J & \multicolumn{2}{|c|}{I-\toh} & I+\toh \\
\hline
\vph  & I=\toh & I>\toh & \\
\hline
\vph A^{(1)}_{J} &
\multicolumn{2}{|c|}{\dst -\frac{I}{2I+1}} &
\dst \frac{I}{2I+1} \\
\hline
\vph A^{(2)}_{J\oh\th} &
\multicolumn{2}{|c|}{\dst \frac{2I}{3(2I+1)}
\left(\frac{2I-1}{I+1}\right)^{1/2}} &
\dst \frac{2I}{3(2I+1)}
\left(\frac{2I+3}{I}\right)^{1/2} \\
\hline
\vph A^{(2)}_{J\th\th} & 0 &
\dst -\frac{I(I+4)}{3(2I+1)(I+1)} &
\dst \frac{I-3}{3(2I+1)} \\
\hline
\end{array}
$$

$$
\renewcommand{\arraystretch}{0}
\begin{array}{|c|c|c|}
\hline
 & \hphantom{\dst -3(2I+1)(I+1)} &
\hphantom{\dst -3(2I+1)(I+1)} \\
\vph J & I-\tth & I+\tth \\
\hline
\vph A^{(2)}_{J\th\th} &
\dst -\frac{I-1}{2I+1} &
\dst \frac{I(I+2)}{(2I+1)(I+1)} \\
\hline
\end{array}
$$
\end{center}
\vspace{\baselineskip}

\newpage

\centerline {\large FIGURE CAPTIONS}
\vspace{\baselineskip}

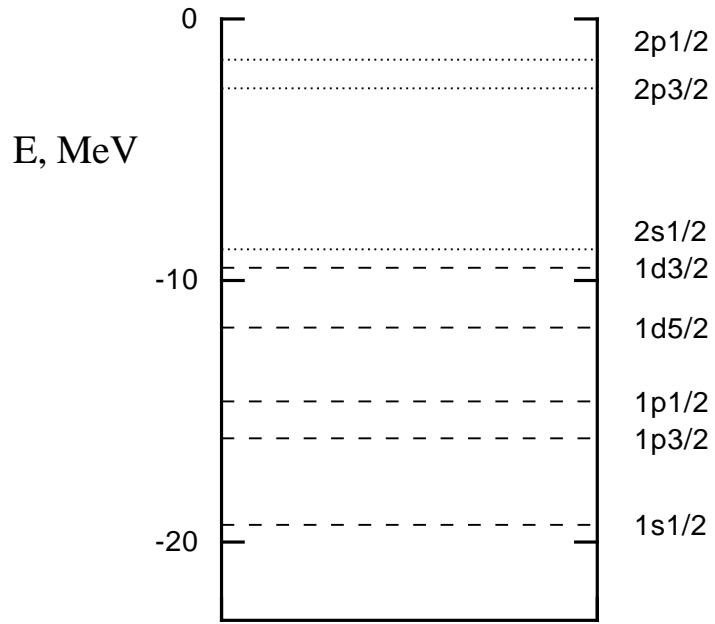
Fig.1. Sequence of levels for the spherical square well of the
radius 6.5~fm. The depth of the well is given by Eq.(\ref{7.1}).
Dashed lines correspond to the bottom levels, on which 18 neutrons
locate. Dotted lines show the free s- and p-levels.
\vspace{\baselineskip}

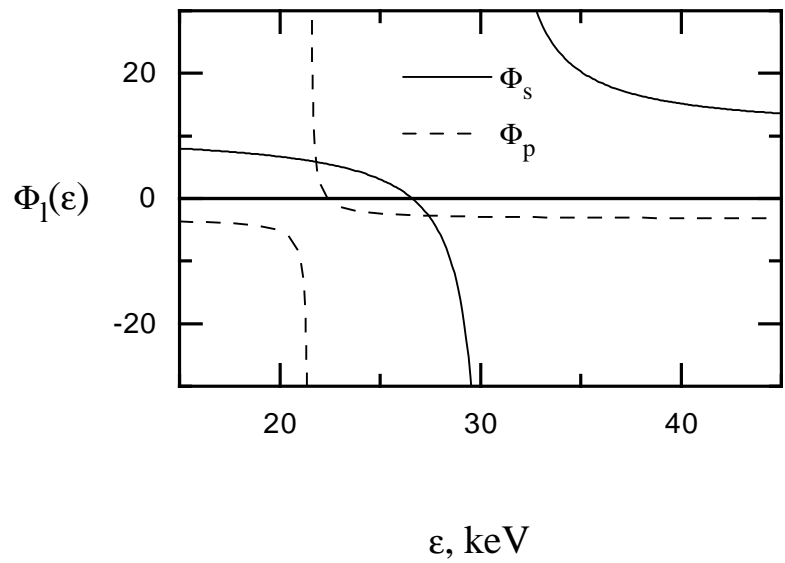
Fig.2. Logarithmic derivatives (\ref{5.35}) of the s- and p-wave
functions of the elastic channel versus the energy.
Solid line --- $\Phi_s(\ve)$, dashed line --- $\Phi_p(\ve)$.
\vspace{\baselineskip}

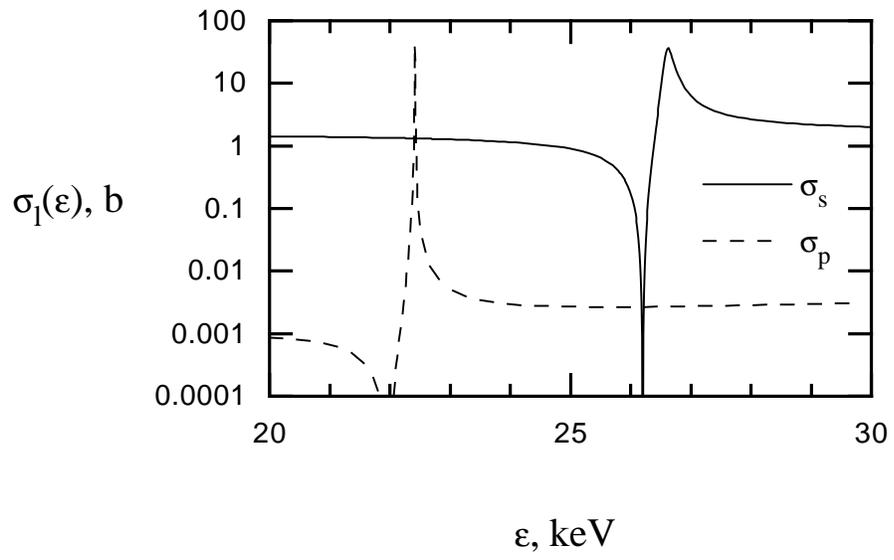
Fig.3. Partial cross sections (\ref{7.4}) of s- and p-wave
scattering versus the energy. Solid line --- $\sigma_s(\ve)$,
dashed line --- $\sigma_p(\ve)$.
\vspace{\baselineskip}

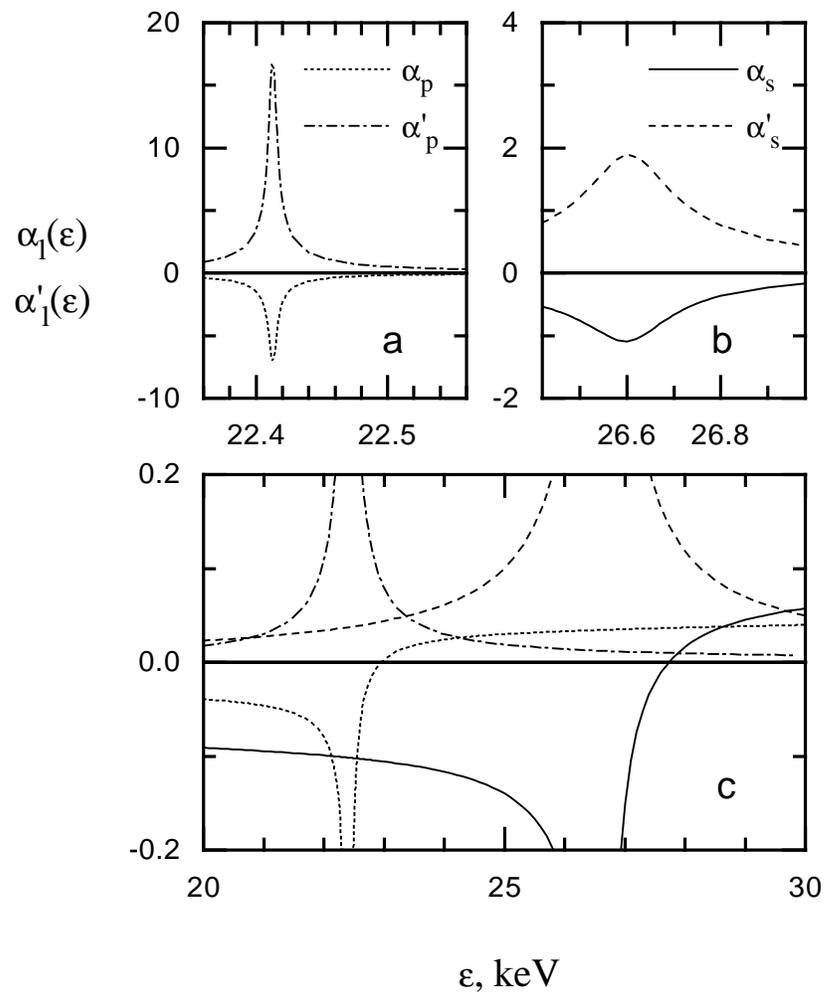
Fig.4. Amplitudes of the s- and p-wave functions of the elastic
channel (\ref{5.8}) near p-wave resonance (a), near s-wave
resonance (b), in a wide range of energy (c).
Solid line --- $\alpha_s(\ve)$, dashed line --- $\alpha'_s(\ve)$,
dotted line --- $\alpha_p(\ve)$,
dash-dotted line --- $\alpha'_p(\ve)$.
\vspace{\baselineskip}

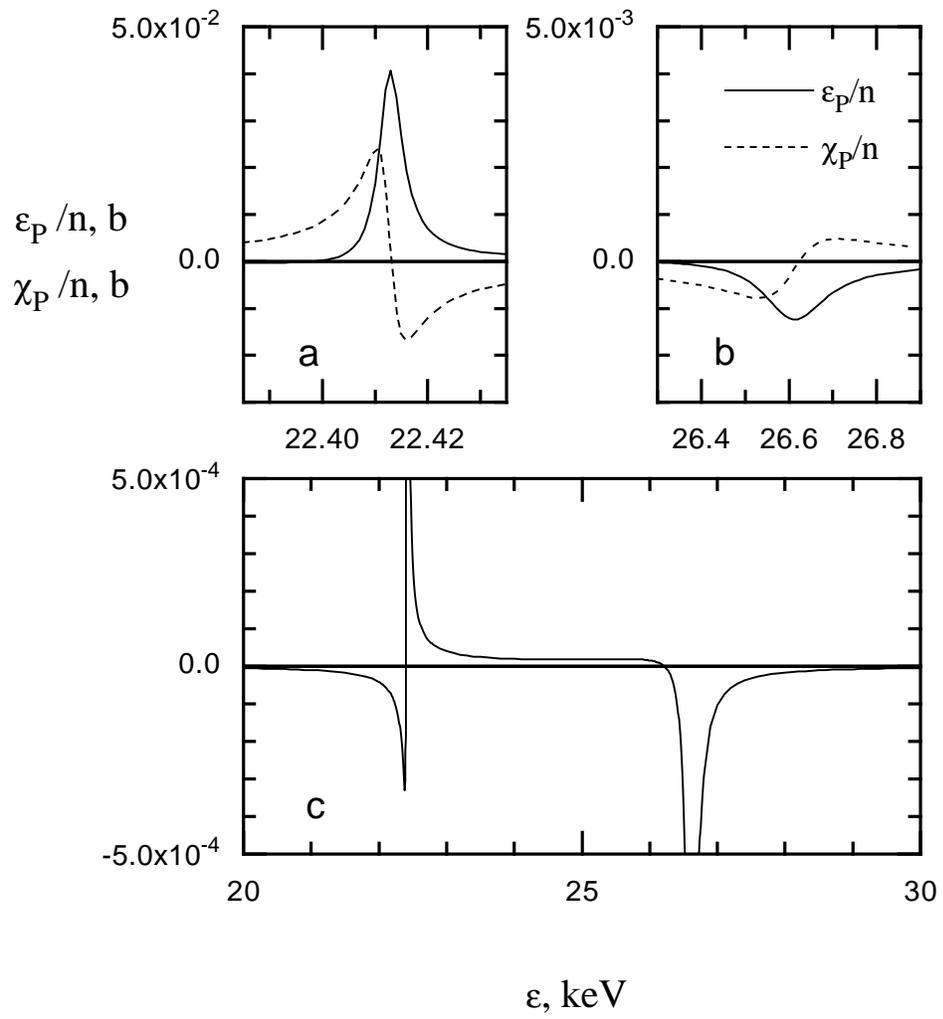
Fig.5. P-noninvariant transmission asymmetry (\ref{7.8}) and angle
of spin rotation (\ref{7.9}) near p-wave resonance (a),
near s-wave resonance (b), in a wide range of energy (c). Solid
line --- $\ep_P/n$, dashed line --- $\chi_P/n$.
\vspace{\baselineskip}

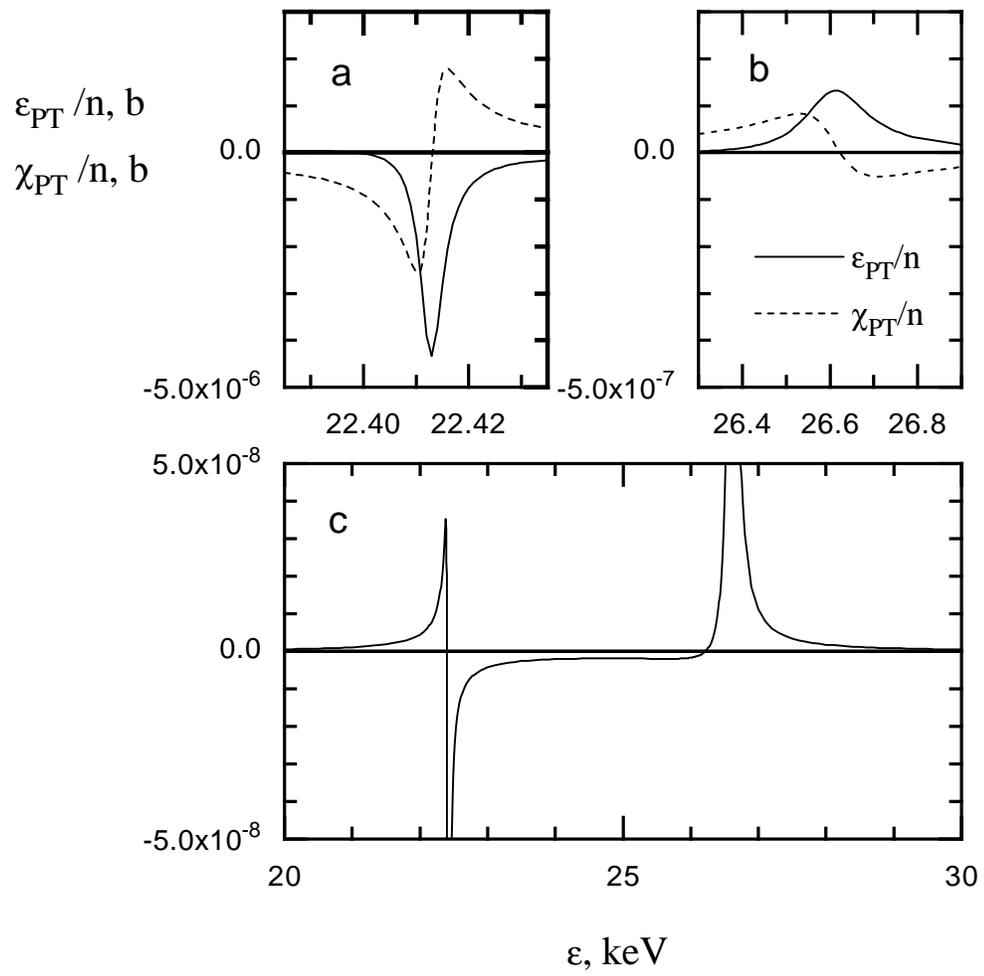
Fig.6. P-,T-noninvariant transmission asymmetry (\ref{7.8}) and
angle of spin rotation (\ref{7.9}) near p-wave resonance (a), near
s-wave resonance (b), in a wide range of energy (c).
Solid line --- $\ep_{PT}/n$, dashed line --- $\chi_{PT}/n$.
\vspace{\baselineskip}

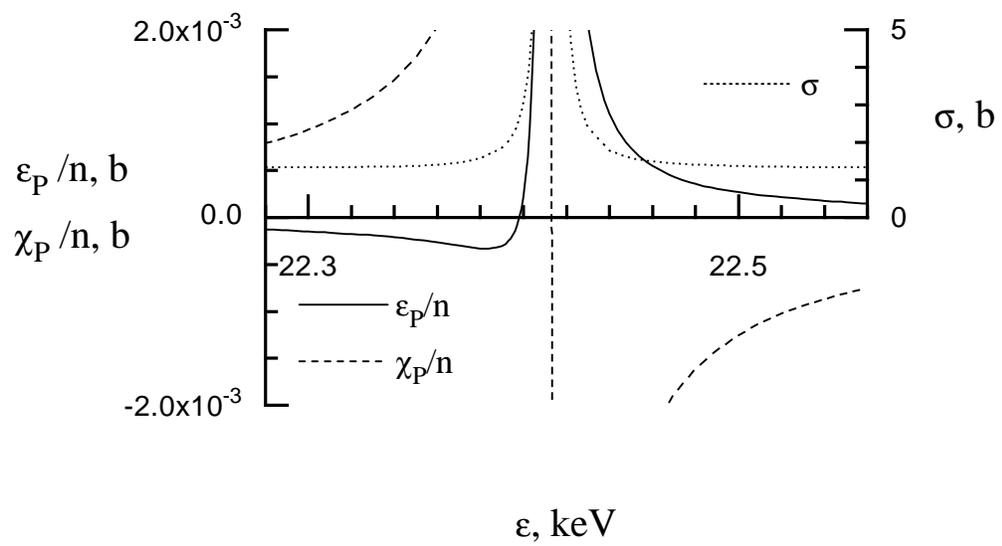
Fig.7. P-noninvariant transmission asymmetry (\ref{7.8}) and angle
of spin rotation (\ref{7.9}), as well as a sum of partial cross
section $\sigma=\sigma_s+\sigma_p$ near p-wave resonance. Solid
line --- $\ep_P/n$, dashed line --- $\chi_P/n$,
dotted line --- $\sigma$.
\vspace{\baselineskip}

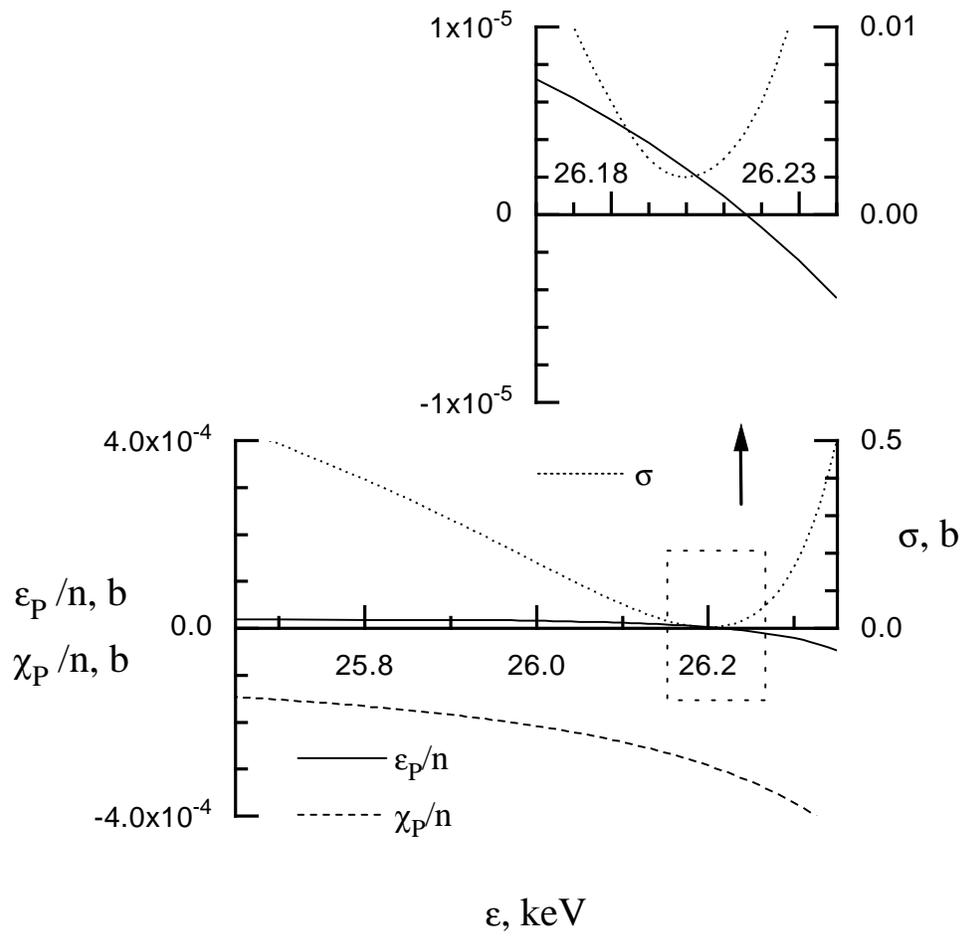
Fig.8. P-noninvariant transmission asymmetry (\ref{7.8}) and angle
of spin rotation (\ref{7.9}), as well as a sum of partial cross
section $\sigma=\sigma_s+\sigma_p$ near an interference minimum
close to the s-wave resonance. Solid line --- $\ep_P/n$,
dashed line --- $\chi_P/n$, dotted line --- $\sigma$.

\end{document}